\documentclass[journal,10pt]{IEEEtran}
\IEEEoverridecommandlockouts
\usepackage{graphicx}
\usepackage{caption}
\usepackage{enumerate}
\usepackage{epsfig}
\usepackage{epstopdf}
\usepackage{cite}
\usepackage{subfigure}
\usepackage{amsmath,amssymb,amsfonts}
\usepackage{algorithmic}
\usepackage{tabularx}
\usepackage{textcomp}
\usepackage{xcolor}
\usepackage{stfloats}
\usepackage{multirow}
\usepackage{url}
\usepackage[ruled]{algorithm2e}
\usepackage{hyperref}
\usepackage{etoolbox,xstring,mfirstuc,textcase}
\usepackage{bm}
\usepackage{amsmath}
\usepackage{amsthm}
\usepackage{stackengine}
\usepackage{cuted}
\usepackage[skip=3pt]{caption}
\usepackage[explicit]{titlesec}

\newtheorem{lemma}{Lemma}
\newtheorem{remark}{Remark}
\def\BibTeX{{\rm B\kern-.05em{\sc i\kern-.025em b}\kern-.08em
    T\kern-.1667em\lower.7ex\hbox{E}\kern-.125emX}}

\titlespacing*{\section}{0pt}{6pt}{0.5pt}  
\titlespacing*{\subsection} {0pt}{4pt}{0.5pt}
\begin{document}

\title{Stacked Intelligent Metasurface-Enhanced \\MIMO OFDM Wideband Communication Systems} 
	\author{Zheao Li, \IEEEmembership{Graduate Student Member, IEEE}, Jiancheng An, \IEEEmembership{Senior Member, IEEE}, and Chau Yuen, \IEEEmembership{Fellow, IEEE}

\thanks{Reference: {\color{blue}{Z. Li, J. An and C. Yuen, “Stacked Intelligent Metasurface-Enhanced MIMO OFDM Wideband Communication Systems," \emph{IEEE Trans. Wireless Commun.}, vol. 25, pp. 9608--9622, 2026, doi: 10.1109/TWC.2025.3636393}}. An earlier version of this paper was presented in part at the IEEE VTS APWCS 2024~\cite{Wideband-SIM}. (\emph{Corresponding author: Chau Yuen}.)}
\thanks{Z. Li, J. An, and C. Yuen are with the School of Electrical and Electronic Engineering, Nanyang Technological University, 639798, Singapore (email: zheao001@e.ntu.edu.sg, \{jiancheng.an, chau.yuen\}@ntu.edu.sg). }
}
\markboth{Accepted to IEEE TRANSACTIONS ON WIRELESS COMMUNICATIONS, doi: 10.1109/TWC.2025.3636393.}%
{Shell \MakeLowercase{\textit{et al.}}: A Sample Article Using IEEEtran.cls for IEEE Journals}

\maketitle

\begin{abstract}
Multiple-input multiple-output (MIMO) orthogonal frequency-division multiplexing (OFDM) systems rely on digital or hybrid digital and analog designs for beamforming against frequency-selective fading, which suffer from high hardware complexity and energy consumption. To address this, this work introduces a fully-analog stacked intelligent metasurfaces (SIM) architecture that directly performs wave-domain beamforming, enabling diagonalization of the end-to-end channel matrix and inherently eliminating inter-antenna interference (IAI) for MIMO OFDM transmission. By leveraging cascaded programmable metasurface layers, the proposed system establishes multiple parallel subchannels, significantly improving multi-carrier transmission efficiency while reducing hardware complexity. To optimize the SIM phase shift matrices, a block coordinate descent and penalty convex-concave procedure (BCD-PCCP) algorithm is developed to iteratively minimize the channel fitting error across subcarriers. Simulation results validate the proposed approach, determining the maximum effective bandwidth and demonstrating substantial performance improvements. Moreover, for a MIMO OFDM system operating at 28 GHz with 16 subcarriers, the proposed SIM configuration method achieves over 300\% enhancement in channel capacity compared to conventional SIM configuration that only accounts for the center frequency.
\end{abstract}
\begin{IEEEkeywords}
Stacked intelligent metasurfaces (SIM), MIMO, OFDM, fully-analog beamforming, wave-based computing.
\end{IEEEkeywords}

\section{Introduction}
~~With the advent of sixth-generation (6G) communication, attention is turning more towards developing innovative technologies that enhance both the spectral and spatial efficiency of wireless networks~\cite{6G1}. Wideband multiple-input multiple-output (MIMO) orthogonal frequency-division multiplexing (OFDM) has emerged as a cornerstone technology, enabling high data rates, robust transmission, and ultra-reliable communication by leveraging spatial multiplexing and multi-carrier transmission~\cite{OFDM,OFDM-2}. In this context, the necessity for more efficient and adaptive antenna array technologies becomes increasingly apparent. These arrays are crucial in supporting the high-frequency wideband communications in 6G networks, which collaboratively promise substantial improvements in data rate and network reliability thanks to the immense bandwidth and the large antenna aperture~\cite{6G2,6GMIMO}. However, the practical implementation of wideband MIMO OFDM systems poses several challenges, particularly in antenna array design, where the need for high-dimensional beamforming and interference mitigation increases system complexity.

Over the past decades, MIMO OFDM antenna architectures have undergone significant evolution with the increase of data throughput and connection density~\cite{OFDM-3}. An issue in wideband MIMO-OFDM systems is the precoding and combining imperfection, which leads to inter-antenna interference (IAI) and limits spatial multiplexing efficiency~\cite{IAI}. Conventional fully-digital antenna arrays mitigate IAI through baseband digital precoding and combining, ensuring independent transmission of spatial streams~\cite{IAI2}. However, this approach requires a large number of RF chains, leading to excessive energy consumption and prohibitive hardware complexity~\cite{fully-digital1,fully-digital2}. To address these challenges, hybrid digital-analog wideband beamforming architectures have been proposed to strike a balance between performance and efficiency by replacing the number of required radio frequency (RF) chains with phase shifters~\cite{HY-1,fully-digital4}. Recently, holographic metasurface antenna (HMA)-based transceiver design has emerged, aiming to optimize the radiation pattern without requiring complex RF structures~\cite{HMIMO,MDR}. As an alternative to traditional antenna arrays, HMAs and reconfigurable intelligent surfaces (RIS) utilize programmable metasurfaces to control signal amplitude and phase~\cite{HMA,HMA1} with greater system flexibility and energy efficiency. Despite these advances, conventional MIMO OFDM paradigms still suffer from high computational complexity, excessive hardware costs, and limited adaptability to wideband signal processing. Additionally, the existing metasurface technologies, for example HMA and RIS, generally adopt a single-layer structure, thus resulting in insufficient capability to fully exploit the EM tuning capability in the wave domain~\cite{SIM-RO}. To bridge these gaps, \textbf{this work proposes a fully-analog stacked intelligent metasurfaces (SIM)-enhanced MIMO OFDM architecture that performs wave-domain beamforming without the power-intensive digital precoding}. By leveraging cascaded programmable metasurface layers, SIM enables direct spatial multiplexing by forming parallel subchannels, inherently suppressing IAI and enhancing spectral efficiency with minimal hardware complexity.

\begin{table*}[h]
\centering
\caption{\centering{\protect\\{\textsc{Comparison of different architectures for MIMO OFDM systems.}}}}\
	\setlength{\tabcolsep}{0.8pt} 
	\renewcommand\arraystretch{1.3} 
\newcommand\xrowht[2][0]{\addstackgap[.5\dimexpr#2\relax]{\vphantom{#1}}}
\label{ofdm}
\begin{tabular}{|m{1.8cm}<{\centering}|m{4cm}<{\centering}|m{4cm}<{\centering}|m{4cm}<{\centering}|m{4cm}<{\centering}|}
\hline
 & {\textbf{Fully-digital antenna array}} & \textbf{Hybrid digital and analog antenna array} & \textbf{HMA-based} & \textbf{SIM-enhanced}  \\ \hline 
\textbf{Schematic} & \includegraphics[width=0.145\textwidth]{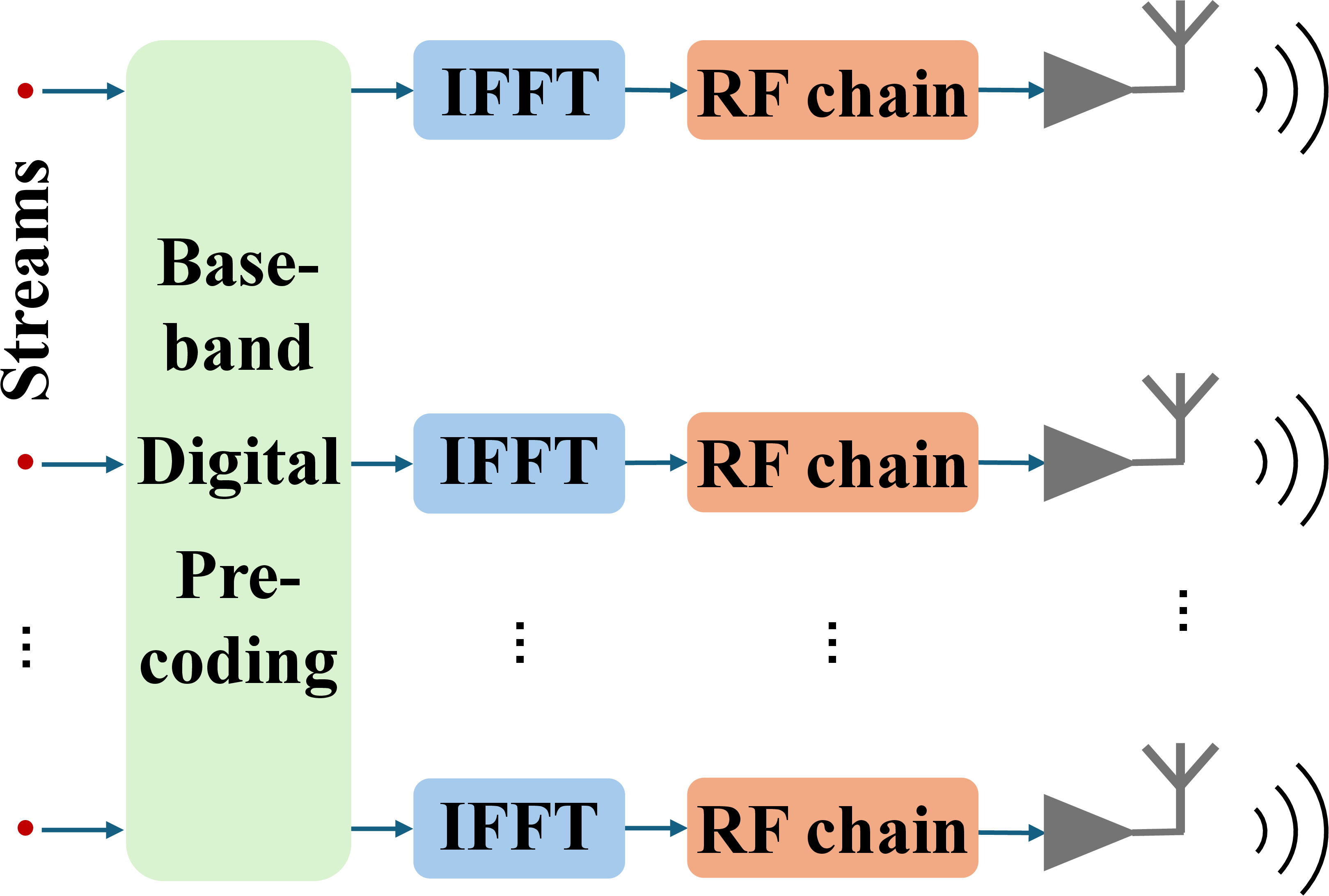} & \includegraphics[width=0.163\textwidth]{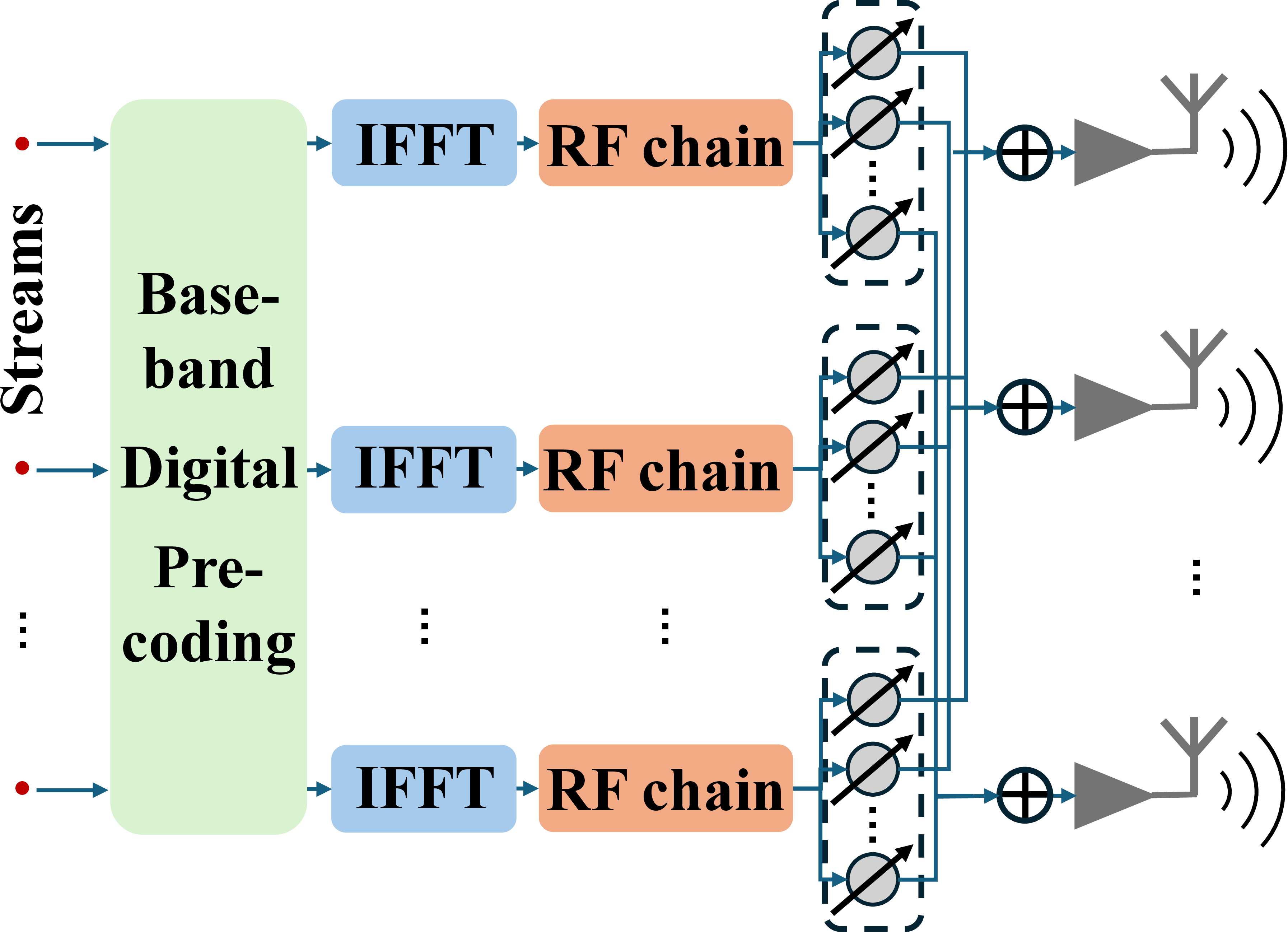} & \includegraphics[width=0.158\textwidth]{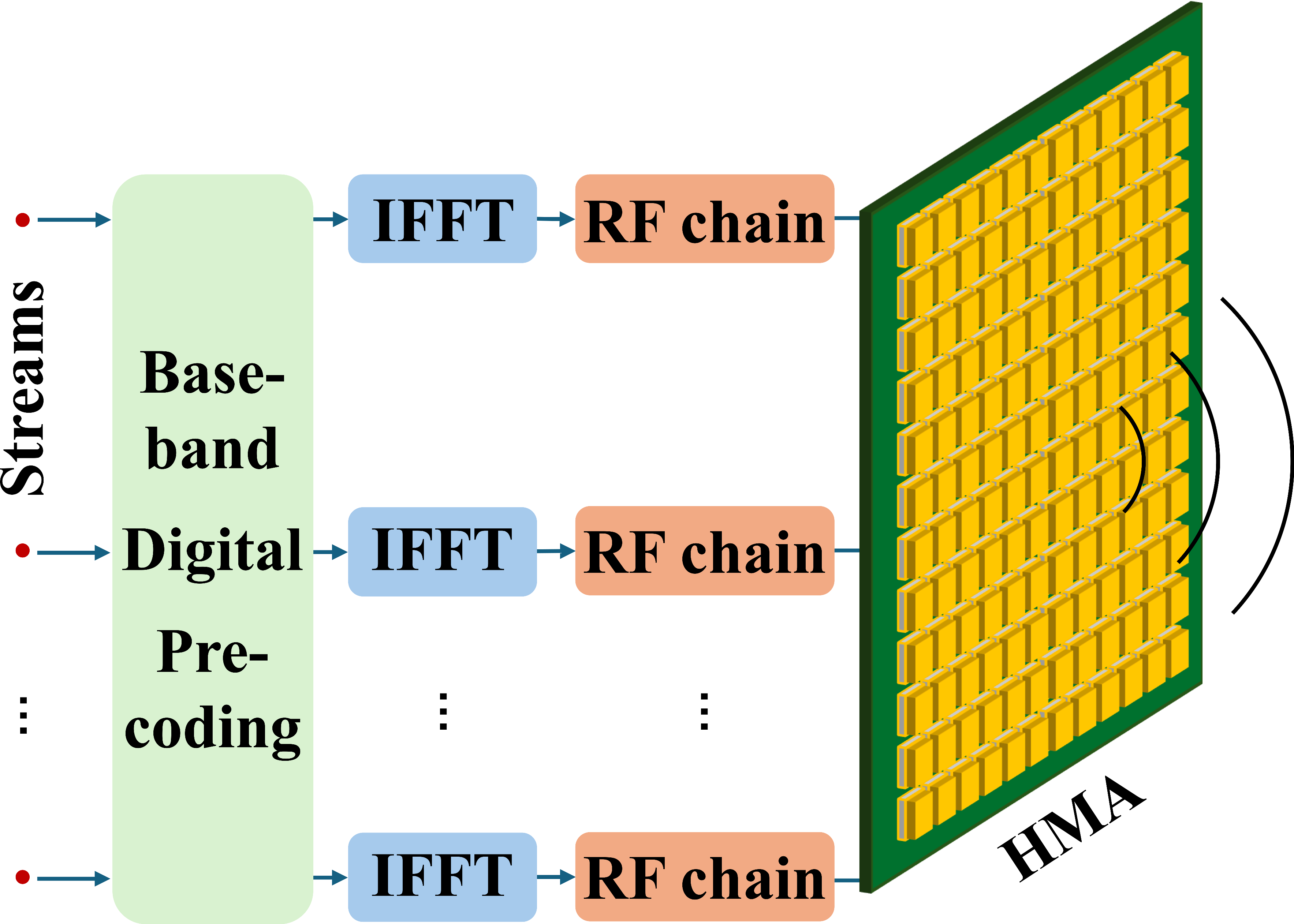} & \includegraphics[width=0.175\textwidth]{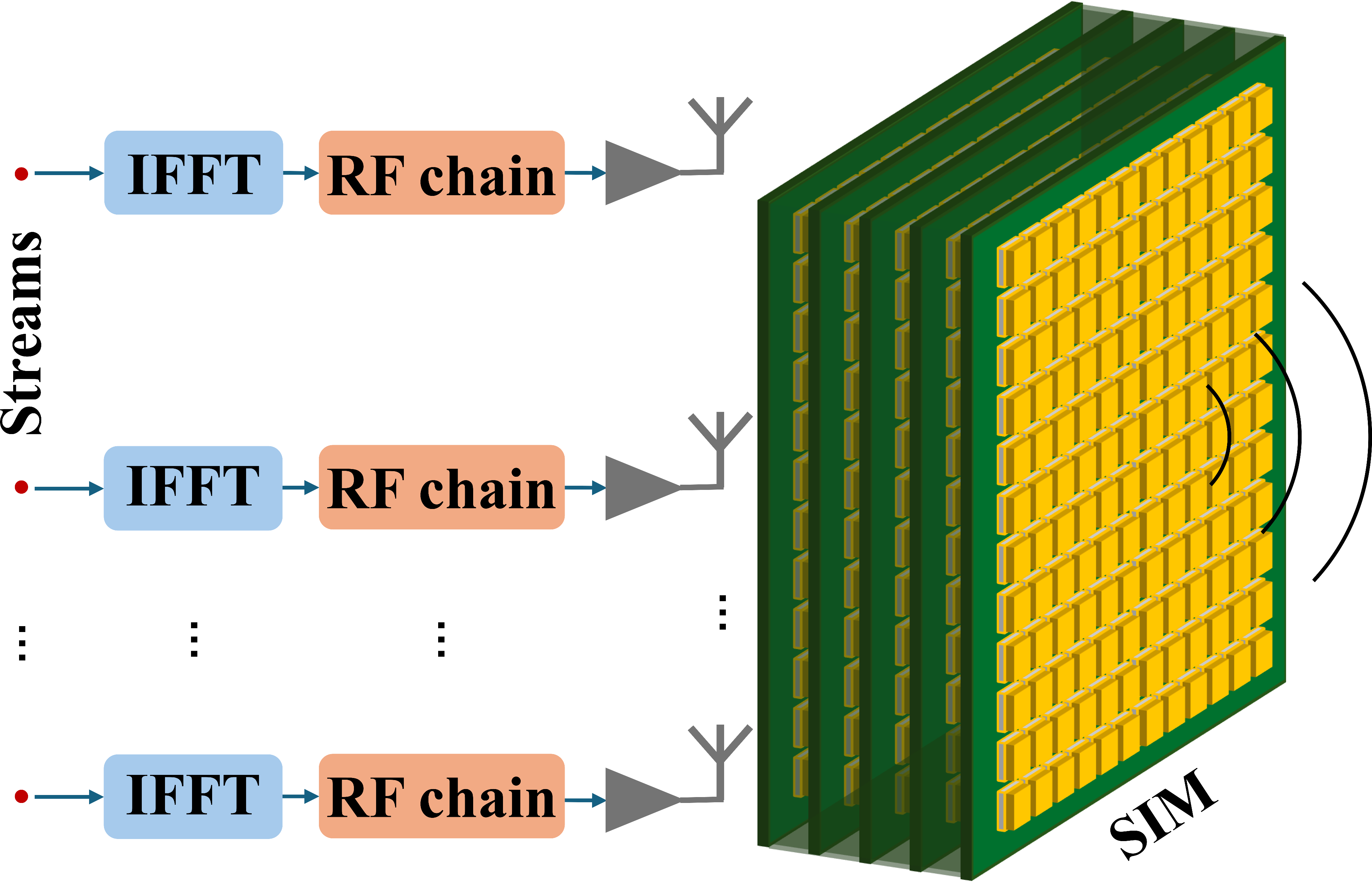} \\ \hline 
\textbf{Complexity} & $O(N_c \cdot N_{TX}\cdot (S+{\rm log}N_c))$ & $O(N_c \cdot N_{RF}\cdot (S+{\rm log}N_c))$ &$O(N_c \cdot N_{RF}\cdot (S+{\rm log}N_c))$ & $O(N_c \cdot S \cdot {\rm log}N_c)$ \\ \hline 
\textbf{Number of RF chains} &$S\le N_{RF}=N_{TX}$,~Large &$S\le N_{RF}\le  N_{TX}$,~Moderate &$S\le N_{RF}\le  N_{TX}$,~Moderate &$N_{RF}=S$,~Small \\ \hline 
\textbf{Hardware cost/Energy consumption} &High &Moderate &Low &Very low \\ \hline 
\textbf{Delay} &High &Moderate &Moderate &Low \\ \hline 
\textbf{Ref.} &\cite{OFDM},~\cite{fully-digital1},~\cite{fully-digital2} &~\cite{HY-1,fully-digital4,OFDM2} &~\cite{HMIMO,MDR,HMA,HMA1}, \cite{OFDM4}, \cite{OFDM3}& \textbf{This work} \\ \hline
\multicolumn{5}{|p{18cm}|}{~~\textbf{Note:} $N_c$ is the number of subcarriers,  $S$ is the number of data streams, $N_{RF}$ is the number of radio frequency (RF) chains, $N_{TX}$ is the number of transmit antennas, and $N_{RX}$ is the number of receive antennas.} \\ \hline
\end{tabular}
\end{table*}

Specifically, a paradigm shift to fully-analog processing has been achieved with the advent of SIM, which is particularly beneficial for simplifying transceiver architectures in wireless communication systems~\cite{SIMHMIMO}. Generally speaking, an SIM is a three-dimensional metasurface architecture that leverages cascaded multi-layer metasurfaces to perform analog computing directly in the electromagnetic (EM) wave domain~\cite{SIM1,SIM2}. In practice, SIMs can function as antenna radomes to adjust the amplitude and phase settings \cite{SIM3}, thus possessing an enhanced capability to redirect signals. Compared to fully-digital and hybrid MIMO OFDM systems, SIM-enhanced fully-analog architecture minimizes the number of RF chains and uses lower-resolution digital-to-analog converters (DAC) and analog-to-digital converters (ADC), significantly reducing hardware costs and improving energy efficiency \cite{SIM-EE}. Because the multi-layer SIM executes MIMO precoding/combining entirely in the wave domain, baseband digital precoders become unnecessary, and the subsequent modulation/detection pipeline is greatly simplified. Moreover, by manipulating propagation across multiple cascaded layers, an SIM attains richer EM wave control ability than HMA or RIS, translating into superior spatial multiplexing and frequency-selective equalization.

The authors of \cite{SIM-HMIMO} and \cite{SIM-HMIMO-MI} demonstrated that SIM is capable of achieving precisely controlled signal transmission and reception for enabling holographic MIMO communications. In \cite{SIM-BF2,SIM-BF3,SIM-BF5}, SIM-based beamforming strategies demonstrated a reduction in energy consumption compared to conventional MIMO approaches. Studies in \cite{SIM3,SIM-BF6, SIM-CE2,SIM-MultiUser1,SIM-MultiUser2,SIM-BF4} also highlighted the advantages of SIM in improving the sum rate, emphasizing its adaptability and effectiveness in handling multi-user communication scenarios. Recent research efforts on direction-of-arrival estimation \cite{SIM-DOA}, channel estimation \cite{SIM-CE1}, satellite communications \cite{SIM-LEO}, communication security \cite{SIM-Secure}, and semantic communications \cite{SIM-SC}, showcased the capabilities of SIM for simplifying system design and enhancing energy efficiency. In addition, SIM’s utility can be extended to cellular and cell-free networks, where \cite{SIM-CF,SIM-CF2,SIM-dp} demonstrated its great promise in managing inter-user interference and resource allocation. Furthermore, SIM can be utilized to enhance precision for wireless sensing tasks \cite{SIM-HW, SIM-S2,SIM-S3}, expanding its applicability in both communication and sensing networks. Recent advances explored reinforcement learning and meta-learning techniques for orchestrating SIM in multi-user systems, further improving resource management and user performance \cite{DRL2,MetaLearning}. Previous studies on SIM have demonstrated its potential for energy-efficient transmission and fully-analog processing in narrowband systems with frequency-flat channels~\cite{SIMsur1,SIMsur2}. However, the application of SIM to wideband MIMO OFDM has not been explored, particularly in addressing IAI and multi-carrier transmission challenges under frequency-selective fading scenarios. To fill this research gap, \textbf{this work explores the potential of SIM for MIMO OFDM systems}. The interaction between EM waves and SIM layers provides a programmable degree of freedom (DoF), where each potential propagation link can be modeled as a tap with adjustable channel impulse responses (CIR)~\cite{Wideband-SIM}. By refining the phase shifts of multiple layers, SIM can be dynamically configured to combat the multi-path effect and enhance multi-carrier transmission. Unlike conventional analog multi-beamforming metasurfaces that are designed primarily for directional radiation control \cite{MBF}, the~proposed~SIM~operates~as~a~fully-analog~spatial signal processor to perform wideband MIMO precoding and combining in the wave domain.

Before going on further, we contrast SIM with other emerging MIMO OFDM architectures, including fully-digital antenna arrays~\cite{OFDM, fully-digital2}, hybrid digital and analog antenna arrays~\cite{fully-digital4, OFDM2}, and HMA-based OFDM systems~\cite{OFDM4, OFDM3} in Table \ref{ofdm}. Specifically, we evaluate the computational complexity, the number of RF chains required, the hardware cost, the energy consumption, and the processing delay of these transmission schemes. The complexity of these systems involves the calculation of the baseband digital precoding matrix and the inverse fast Fourier transform (IFFT) for OFDM signal generation. Specifically, the baseband digital precoding matrix maps $S$-dimensional data streams to $N_{RF}$-dimensional RF chains or $N_{TX}$-dimensional transmit antennas, enabling spatial multiplexing and beamforming. For each antenna, the IFFT operation further transforms frequency-domain signals across $N_c$ subcarriers into time-domain signals. Fully-digital systems, while offering high flexibility, suffer from extremely high complexity $O(N_c \cdot N_{TX}\cdot (S+{\rm log}N_c))$, and require a large number of RF chains $N_{RF}=N_{TX}$, leading to significant hardware costs, energy consumption, and processing delays. Hybrid digital and analog architectures, relying on phase shifters and HMA, reduce the number of RF chains to $N_{RF}\le N_{TX}$ and, accordingly, the computational burden is reduced to $O(N_c \cdot N_{RF}\cdot (S+{\rm log}N_c))$. {In contrast, the proposed SIM-enhanced MIMO OFDM system introduces a transformative approach by leveraging fully-analog beamforming in the wave domain, achieving $N_{RF}=N_{TX}=S$. This wave-based computing paradigm provides a highly efficient and scalable design, leading to significantly reduced complexity $O(N_c \cdot S \cdot {\rm log}N_c)$, hardware cost, energy consumption, and processing delay. Additionally, the SIM's wave-based computing capability can achieve the precision and flexibility of digital systems.} 

The main contributions and innovations of this research are described as follows.
\begin{itemize}
\item A novel SIM-enhanced MIMO OFDM architecture is proposed, where fully-analog beamforming is directly performed in the wave domain. By leveraging cascaded programmable metasurface layers, the system naturally forms parallel subchannels, effectively mitigating IAI and enhancing spatial multiplexing. Additionally, the programmable multi-path propagation within SIM improves wideband transmission by mitigating deep fading effects, ensuring robustness against frequency-selective scenarios. 

\item  An optimization problem is formulated to approximate an end-to-end diagonal channel matrix across multiple subcarriers by optimizing the phase shifts of the metasurface layers. This formulation enables each spatial stream to be independently transmitted and received from the corresponding antenna, creating interference-free parallel subchannels across the system bandwidth. Unlike the SIM configuration considering only the center frequency, this formulation aims to accommodate multi-carrier transmission, addressing the frequency-selective fading prevalent in wideband channels.

\item A block coordinate descent-penalty convex concave procedure (BCD-PCCP) algorithm is proposed to address the non-convex optimization problem arising from coupled variables and unit-modulus constraints. Comprehensive theoretical analyses confirm its computational efficiency and superior performance in achieving higher channel capacity and system gain compared to existing methods.

\item  A thorough investigation into the optimal SIM configurations and the impact of key system parameters is conducted, including effective bandwidth, subcarrier spacing, and the number of subcarriers. Numerical results reveal that the SIM-enhanced MIMO OFDM system achieves better adaptability to frequency-selective channels and significantly outperforms conventional SIM configurations that focus solely on the center frequency. Simulation results also validate that the proposed system achieves substantial performance improvements, including higher channel capacity and better wideband channel fitting, over single-layer metasurface designs.
\end{itemize}

The remainder of this paper is organized as follows: Section~\ref{Section2} details the proposed SIM-enhanced MIMO OFDM communication system. The formulation of the channel fitting problem and the developed BCD-PCCP solution strategies are then provided in Section~\ref{Section3}. In Section~\ref{Section4}, the convergence and complexity analysis of the system model is presented. The experimental results and analysis are provided in Section~\ref{Section5}. Finally, conclusions are drawn in Section~\ref{Section6}.

\textbf{Notations:}
In this paper, bold lowercase and uppercase letters are used to denote vectors and matrices, respectively; $ (\mathbf{A})^* $, $ (\mathbf{A})^T $, and $ (\mathbf{A})^H $ represent the conjugate, transpose, and Hermitian transpose of matrix $\mathbf{A}$, respectively; $ |c| $, $ \Re(c) $, and $ \Im(c) $ refer to the magnitude, real part, and imaginary part, respectively, of a complex number $ c $; $ \|\mathbf{A}\|_F $ denotes the Frobenius norm; $ \mathbb{E}(\mathbf{A}) $ stands for the expectation operator; $ \text{diag}(\mathbf{v}) $ produces a diagonal matrix with the elements of vector $ \mathbf{v} $ on the main diagonal; $ \text{vec}(\mathbf{A}) $ denotes the vectorization of a matrix $ \mathbf{A} $; $ \mathbf{A}_{a:b, :} $ and $ \mathbf{A}_{:, c:d} $ represent the submatrices constructed by extracting rows $ a $ to $ b $ and columns $ c $ to $ d $ from matrix $ \mathbf{A} $, respectively; $\rm Tr(\mathbf{A})$ represents the trace of a matrix $ \mathbf{A} $; The Kronecker product and Hadamard product between matrices $\mathbf{A}$ and $\mathbf{B}$ are denoted by $\mathbf{A} \otimes\mathbf{B}$ and $\mathbf{A}\odot\mathbf{B}$, respectively; $ \mathbb{C}^{x \times y} $ represents the space of $ x \times y $ complex-valued matrices; $ {\partial f}/{\partial x} $ denotes the partial derivative of a function $ f $ with respect to (w.r.t.) the variable $ x $.

\section{The Proposed SIM-enhanced MIMO OFDM Wideband Communication System Model}
~~In this section, the architecture and operational principles of the proposed SIM-enhanced MIMO OFDM communication system are introduced.

\label{Section2}
\subsection{SIM-enhanced MIMO OFDM Transceiver Architecture}
\begin{figure*}
	\centerline{\includegraphics[width=1\textwidth]{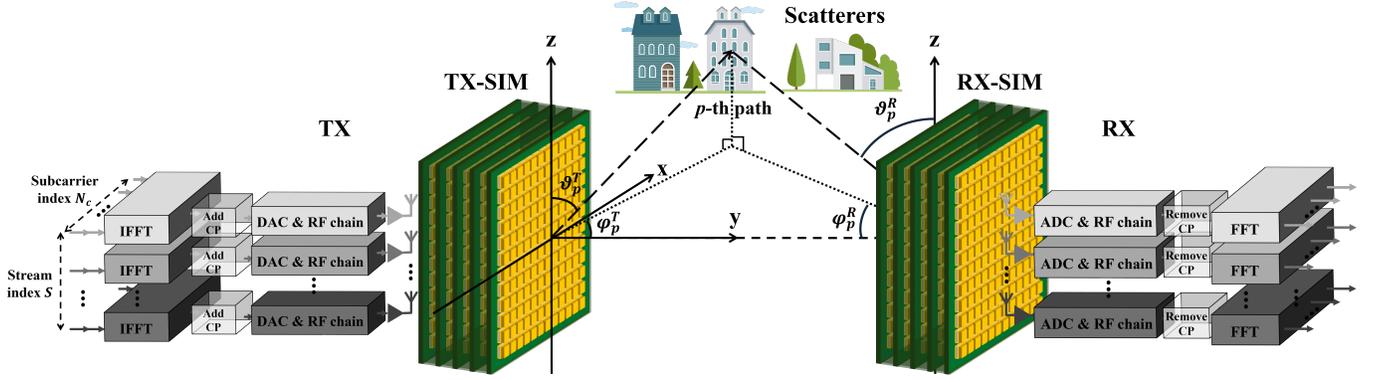}}
	\caption{\centering{The SIM-enhanced wideband MIMO OFDM communication system.}}
	\label{SIM-channel}
\end{figure*}

~~The proposed SIM-enhanced MIMO OFDM architecture is illustrated in Fig.~\ref{SIM-channel}. We consider a wideband communication system over $N_c$ subcarriers, where two SIMs are utilized to transmit and receive $S$ independent data streams on each subcarrier. By performing beamforming directly in the wave domain~\cite{SIM-HMIMO}, the system eliminates the need for baseband digital precoding and combining. This streamlined architecture minimizes hardware complexity, energy consumption, and computational overhead, as it requires fewer RF chains. {In this design, the data stream associated with each transmit antenna is independently processed and received at a corresponding receive antenna, because multiple parallel subchannels are established in the physical space. Therefore, unlike conventional MIMO architectures, where the number of RF chains $N_{RF}$ scales with the number of transmit antennas $N_{TX}$, the proposed system achieves $N_{RF} = N_{TX} = S$, which ensures that each spatial stream is physically realized and independently excitable. For the sake of simplicity, we set the receive antenna number equal to the transmit antenna number $N_{RX}= N_{TX}$ to form a square end-to-end channel and to demonstrate the effect of the SIM on spatial multiplexing.} This fully-analog beamforming approach effectively suppresses IAI, thereby enhancing signal quality and improving overall system robustness.

Specifically, at the transmitter (TX), $S$ data streams are first modulated in the frequency domain. The data stream over $N_c$ subcarriers for each transmit antenna undergoes an IFFT operation to convert frequency-domain signals into time-domain OFDM signals. A cyclic prefix (CP) is then added to combat inter-symbol interference caused by multi-path propagation. The processed OFDM signals are passed through $S$ DACs and RF chains, followed by the wave-domain beamforming implemented through the TX-SIM structure. At the receiver (RX), the wideband signals are received through the RX-SIM, which performs wave-domain processing to mitigate wideband channel impairment and IAI. After passing through $S$ RF chains and ADCs and removing the CP, the time-domain signals are subsequently transformed back into the frequency domain via fast Fourier transform (FFT) operations to recover the original $N_c$ symbols for each data stream. This end-to-end process ensures efficient wideband data transmission and reception while leveraging the SIM's ability to reduce complexity and compensate for channel distortions.

\begin{figure}
	\centerline{\includegraphics[width=0.5\textwidth]{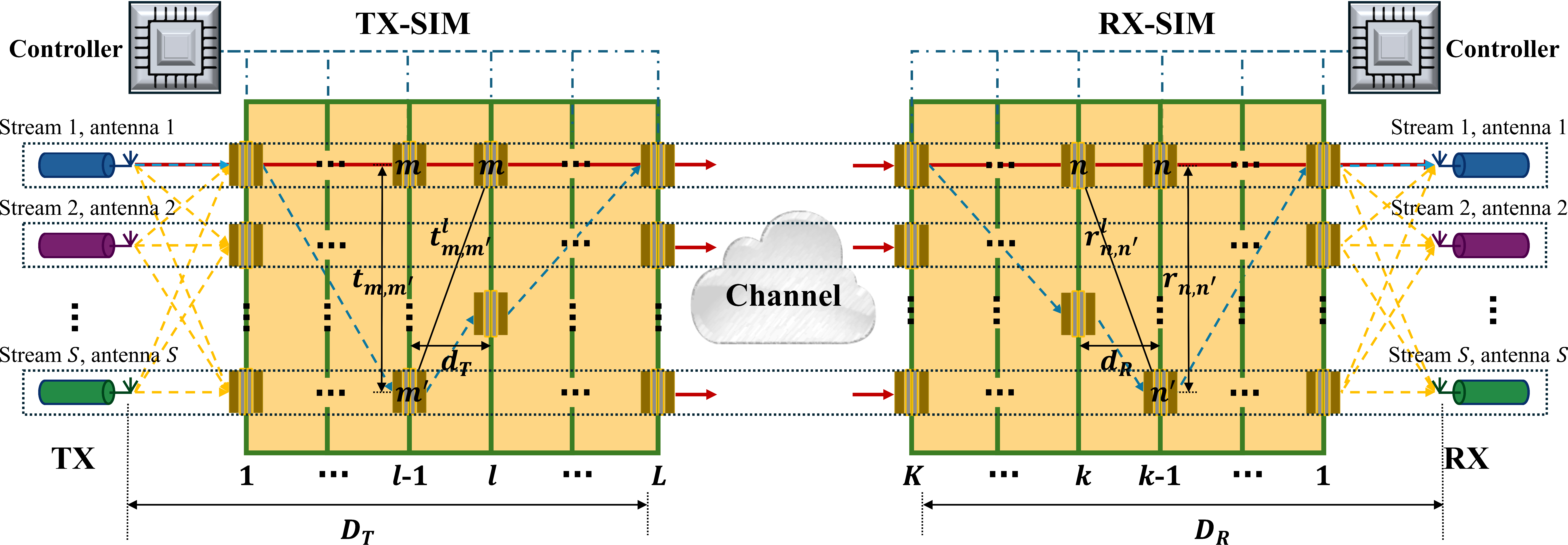}}
	\caption{\centering{The architecture of the SIM.}}
	\label{SIM}
\end{figure}

More specifically, the TX-SIM and RX-SIM structures are built by cascading multiple intelligent metasurfaces that can dynamically interact with incident EM wavefronts in the fully-analog wave domain for efficient signal processing. The system involves $L$ metasurface layers at the TX-SIM and $K$ layers at the RX-SIM, with both structures connected to intelligent controllers, as illustrated in Fig.~\ref{SIM}. Each metasurface layer in TX-SIM contains $M$ meta-atoms and $N$ in RX-SIM. For each layer of the TX-SIM, the spacing between the $m$-th and $m'$-th meta-atoms is represented as $t_{m, m'}$, while the inter-layer spacing between the $m$-th meta-atom of the $l$-th layer and the $m'$-th meta-atom of the ($l$-1)-th layer is denoted as $t_{m, m'}^l$. Similarly, in the RX-SIM, the intra-layer spacing between the $n$-th and $n'$-th meta-atoms is $r_{n, n'}$, and the inter-layer spacing between the $n$-th meta-atom of the $k$-th layer and the $n'$-th meta-atom of the ($k$-1)-th layer is $r_{n, n'}^k$. The inter-layer distances in the TX-SIM and RX-SIM are $d_T$ and $d_R$, respectively, with total thicknesses denoted as $D_T$ and $D_R$. This compact and meticulously designed architecture facilitates precise control over EM wave properties, enabling efficient wideband signal processing with enhanced performance and reduced complexity.

According to the Rayleigh-Sommerfeld diffraction theory~\cite{RS}, the wideband transmission coefficients of the TX-SIM and RX-SIM are expressed as:
\begin{align}
w_{m, m'}^l(f_i)&=\frac{S_Td_T}{{(t_{m, m'}^l)}^2} (\frac{1}{2\pi t_{m, m'}^l}-j\tfrac{f_i}{c})e^{j2\pi t_{m, m'}^lf_i/c}, \label{co1} \\
u_{n, n'}^k(f_i)&=\frac{S_Rd_R}{{(r_{n, n'}^k)}^2} (\frac{1}{2\pi r_{n, n'}^k}-j\tfrac{f_i}{c })e^{j2\pi r_{n, n'}^kf_i/c},\label{co2} 
\end{align}
where $S_T$ and $S_R$ are the areas of each meta-atom in the TX-SIM and RX-SIM, respectively, $f_i$ denotes the frequency of the $i$-th subcarrier, and $c$ is the speed of light. 

Each meta-atom in the SIM adjusts its transmission coefficient by imposing a phase shift, represented by $\phi_m^l=e^{j\theta_m^l}$ for the $m$-th meta-atom on the $l$-th TX-SIM layer and $\psi_n^k=e^{j\zeta_n^k}$ for the $n$-th meta-atom on the $k$-th RX-SIM layer. {According to \cite{continuous}, continuously adjustable phase shifts in the interval between 0 and $2\pi$ are considered for the transmission coefficients associated with each meta-atom.} These phase shift matrices of the $l$-th TX-SIM layer and the $k$-th RX-SIM layer are denoted by $\mathbf{\Phi}^l={\rm diag}([\phi_1^l,  \phi_2^l, \ldots, \phi_M^l]^T)\in \mathbb{C}^{M\times M}$ and $\mathbf{\Psi}^k={\rm diag}([\psi_1^k,  \psi_2^k, \ldots, \psi_N^k]^T)\in \mathbb{C}^{N\times N}$, respectively. The cumulative effect of signal propagation through the layers of the TX-SIM and RX-SIM is characterized by the transmission functions $\mathbf{P}$ and $\mathbf{Q}$, respectively. For the TX-SIM, the transmission function of the $i$-th subcarrier is expressed as:
\begin{equation}
\label{Pi}
\mathbf{P}_i=\mathbf{P}(f_i)=\mathbf{\Phi}^L\mathbf{W}^L_i...\mathbf{\Phi}^2\mathbf{W}^2_i\mathbf{\Phi}^1\mathbf{W}^1_i\in \mathbb{C}^{M\times S},
\end{equation}
where $[\mathbf{W}^l_i]_{m, m'}= w_{m, m'}^l(f_i),~l=2, \ldots, L$ denotes the wideband transmission matrix between the ($l$-$1$)-th layer and the $l$-th layer of the TX-SIM and $\mathbf{W}^1_i\in \mathbb{C}^{M\times S}$ describes the transmission from the transmit antenna array to the first layer of the TX-SIM.

Similarly, the transmission function for the RX-SIM of the $i$-th subcarrier is given by:
\begin{equation}
\label{Qi}
\mathbf{Q}_i=\mathbf{Q}(f_i)=\mathbf{U}^1_i\mathbf{\Psi}^1\mathbf{U}^2_i\mathbf{\Psi}^2...\mathbf{U}^K_i\mathbf{\Psi}^K\in \mathbb{C}^{S\times N},
\end{equation}
where $[\mathbf{U}^k_i]_{n, n'}= u_{n, n'}^l(f_i),~k=2, \ldots, K$ represents the wideband transmission matrix between the ($k$-$1$)-th layer and the $k$-th layer of the RX-SIM and $\mathbf{U}^1_i\in \mathbb{C}^{S\times N}$ defines the transmission from the first layer of the RX-SIM to the receive antenna array.

\begin{remark}
\textnormal{The interaction of waves through multiple layers of SIM can be analogized to multi-path signal propagation, where the transmission via each meta-atom can be modeled as a tap with adjustable CIR. As shown in Fig.~\ref{SIM}, the multi-layered structure of the SIM provides flexible control over the EM wave propagation through metasurface layers. By dynamically adjusting the phase shift matrices of each metasurface layer, this dynamic wave-domain manipulation in \eqref{Pi} and \eqref{Qi} allows the SIM to achieve dual functions: 
\begin{enumerate}
\item Effectively compensate for frequency-selective fading caused by multi-path effects in wideband channels.
\item Construct $S$ parallel interference-free transmission subchannels over $N_c$ subcarriers.
\end{enumerate}
}
\end{remark}

\begin{remark}
\textnormal{Note that existing SIM prototypes can be categorized as three hardware classes~\cite{SIM-HW}: (i) H-S—static SIMs without control circuits; (ii) H-P—programmable yet passive SIMs that adjust only the reactive component of the surface impedance via PIN or other microelectromechanical system designs; and (iii) H-A—programmable and active SIMs that embed gain elements for joint phase-amplitude control. In this paper we adopt the H-P transmissive regime to develop a fully-analog, wave-domain beamforming methodology that minimizes the number of RF chains and enables low-resolution DAC/ADC operation, thereby reducing hardware cost and energy consumption. A notable H-A variant is the parity-time (PT) symmetric metasurface, where balanced loss–gain unit cells enable simultaneous phase–amplitude control and coherent-perfect absorption~\cite{PTS}. The integration of PT-symmetric metasurfaces into the stacked architecture may further enhance the functionality of SIM, while the detailed algorithm design requires further investigation.} 
\end{remark}

\begin{remark}
\textnormal{In our design and prior transmissive SIM prototypes~\cite{SIM1}, unidirectional meta-atoms are designed to ensure forward-only power flow, thereby preventing loopback coupling and self-exciting oscillations. Consequently, the power transfer within the stacked structure is predominantly forward. Each metasurface layer is enclosed within a metallic frame surrounded by microwave-absorbing materials to damp edge/side diffraction, while per-layer reflections decay exponentially at higher orders. In addition, a thin absorbing/low-reflection coating is typically applied behind the stack to further suppress any residual return. Therefore, the forward-propagation assumption adopted in our model is physically well-justified. Furthermore, one may consider exploiting controlled inter-layer feedback instead of suppressing it so that a SIM can be utilized as a wave-domain recursive network in future work.}
\end{remark}

\begin{remark}
\textnormal{The SIM is assembled from patterned transmissive panels separated by precision spacers. Inter-layer registration uses fiducials and alignment holes/dowel pins integrated into the panels and spacer frame. Standard printed circuit board (PCB) lamination or clamped frames typically provide $\mathcal{O}(10^2\,\mu\text{m})$ repeatability at mmWave apertures, which is commensurate with the slowly varying (supercell-scale) phase profiles optimized in this work. In addition, update either the transmission coefficients in \eqref{co1} and \eqref{co2} by minimizing the field mismatch (e.g., least squares with back-propagation through the stack), whose procedure compensates for residual misalignment and modeling errors.}
\end{remark}

\subsection{Wideband MIMO Channel Model}
~~In the proposed SIM-enhanced MIMO OFDM system, a geometric-based multi-path channel model is adopted to accurately capture the spatial and frequency-selective characteristics of wideband propagation. The channel accounts for $P_s$ scatterers located in the far field of the TX-SIM and RX-SIM, where each scatterer contributes to a corresponding path in the propagation environment. Let $p = 0$ denote the line-of-sight (LoS) path and $p=1,\ldots,P_s$ denote non-LoS (NLoS) paths. The complex gain and delay associated with the $p$-th path are denoted as $g_p$ and $\tau_p$, respectively. Although the transmission signals between TX and RX are time-domain signals, the wideband channel model is often represented in the frequency-domain form to simplify the IFFT-FFT progress in the beamspace \cite{Fchannel}. For the $i$-th subcarrier with frequency $f_i$, the MIMO channel $\mathbf{G}(f_i)\in \mathbb{C}^{N\times M}$ between the TX-SIM and RX-SIM is expressed as:
\begin{align}
\label{channel}
\mathbf{G}_i=\mathbf{G}(f_i)=  \sum_{p=0}^{P_s}g_{p}(f_i)e^{-j2\pi f_i\tau_p}\bm{\alpha}_p^r(f_i){\bm{\alpha}_p^t(f_i)}^H, 
\end{align}
where $\bm{\alpha}_p^t(f_i) \in \mathbb{C}^{M\times1}$ and $\bm{\alpha}_p^r(f_i) \in \mathbb{C}^{N\times1}$ are the steering vectors of the $p$-th path w.r.t. the TX-SIM and the RX-SIM of the $i$-th subcarrier, respectively.

Without loss of generality, we assume that the TX-SIM and RX-SIM are vertically aligned in a 3D Cartesian coordinate system. The center of the TX-SIM is located at the origin (0, 0, 0). As illustrated in Fig.~\ref{SIM-channel}, the orientation of the $p$-th scatterer relative to the TX-SIM is characterized by the physical elevation angle $\vartheta_p^t \in [0,\pi)$ and the azimuth angle $\varphi_p^t \in [-\pi/2,\pi/2]$, while its orientation relative to the RX-SIM is described by $\vartheta_p^r \in [0,\pi)$ and $\varphi_p^r \in [-\pi/2,\pi/2]$. Each metasurface layer in the TX-SIM comprises $M = M_x \times M_z$ meta-atoms, where $M_x$ and $M_z$ denote the number of meta-atoms along the $x$-axis and $z$-axis, respectively. Similarly, each layer in the RX-SIM consists of $N = N_x \times N_z$ meta-atoms, with $N_x$ and $N_z$ representing the number of meta-atoms along the $x$-axis and $z$-axis, respectively. The spacing between adjacent meta-atoms in the TX-SIM is denoted as $r_T$, and the spacing in the RX-SIM is denoted as $r_R$.

For simplicity, the azimuth and elevation components of the $p$-th scatterer relative to the meta-atom positions in the TX-SIM and RX-SIM are projected into the spatial frequency domain. Specifically, the electrical angles $\psi_p^{t_x}$ and $\psi_p^{t_z}$ are defined for the $x$ and $z$ dimensions of the TX-SIM, respectively. Similarly, the electrical angles $\psi_p^{r_x}$, $\psi_p^{r_z}$ are defined for the $x$ and $z$ dimensions of the RX-SIM, respectively. The electrical angles can be expressed by \cite{UPA}
\begin{align}
\psi_p^{t_x}(f_i)& = k_ir_{T}\sin(\vartheta_p^{t})\sin(\varphi_p^{t}),\\
\psi_p^{t_z}(f_i)& = k_ir_{T}\cos(\vartheta_p^{t}),\\
\psi_p^{r_x}(f_i)& = k_ir_{R}\sin(\vartheta_p^{r})\sin(\varphi_p^{r}),\\
\psi_p^{r_z}(f_i)& = k_ir_{R}\cos(\vartheta_p^{r}),
\end{align}
where $k_i=2\pi f_i/c$ represents the wavenumber of the $i$-th subcarrier.

Steering vectors of TX-SIM $\bm{\alpha}_p^t(f_i)$ and RX-SIM $\bm{\alpha}_p^r(f_i)$ can be written as 
\begin{align}
\bm{\alpha}_p^{t}(f_i) &= \bm{\alpha}_p^{t_x}(\psi_p^{t_x}(f_i)) \otimes \bm{\alpha}_p^{t_z}(\psi_p^{t_z}(f_i)),\\
\bm{\alpha}_p^{r}(f_i) &= \bm{\alpha}_p^{r_x}(\psi_p^{r_x}(f_i)) \otimes \bm{\alpha}_p^{r_z}(\psi_p^{r_z}(f_i)),
\end{align}
where $\bm{\alpha}_p^{t_x}(\psi_p^{t_x}(f_i))\in \mathbb{C}^{M_x\times1}$ and $\bm{\alpha}_p^{t_z}(\psi_p^{t_z}(f_i))\in \mathbb{C}^{M_z\times1}$,  are defined as follows:
\begin{equation}
[\bm{\alpha}_p^{t_x}(\psi_p^{t_x}(f_i))] _{m_x} \triangleq  e^{j(m_x-1)\psi_p^{t_x}(f_i)},~m_x = 1,\ldots, M_x,
\end{equation}
\begin{equation}
[\bm{\alpha}_p^{t_z}(\psi_p^{t_z}(f_i))] _{m_z} \triangleq  e^{j(m_z-1)\psi_p^{t_z}(f_i)},~m_z = 1,\ldots, M_z,
\end{equation}
where $\bm{\alpha}_p^{r_x}(\psi_p^{r_x}(f_i))\in \mathbb{C}^{N_x\times1}$ and $\bm{\alpha}_p^{r_z}(\psi_p^{r_z}(f_i))\in \mathbb{C}^{N_z\times1}$ of the RX-SIM are expressed in the same way.

\subsection{SIM-Enhanced MIMO OFDM System Model}
~~For the $i$-th subcarrier, given the wireless channel $\mathbf{G}_i$ and a fixed number of data streams $S$, the optimal transmission strategy for the proposed SIM-enhanced MIMO OFDM system can be determined using singular value decomposition (SVD). The wideband MIMO channel matrix $\mathbf{G}_i$ is decomposed as:
\begin{equation}
\mathbf{G}_i=\mathbf{E}_i\mathbf{\Lambda}_i\mathbf{F}_i^H,
\end{equation}
where $\mathbf{\Lambda}_i$ is a diagonal matrix containing the singular values $[{\lambda}_i]_1, [{\lambda}_i]_2, \ldots, [{\lambda}_i]_{\min(M, N)}$ of the $i$-th subcarrier sorted in non-increasing order. $\mathbf{E}_i$ and $\mathbf{F}_i$ are unitary matrices that captures the left and right singular vectors of $\mathbf{G}_{i}$, respectively. 

To achieve fully-analog spatial multiplexing, we directly apply the precoder and combiner in the frequency domain, removing partial digital baseband processing: $\mathbf{P}_i = [\mathbf{F}_i]_{:, 1:S} \in \mathbb{C}^{M \times S}$ and $\mathbf{Q}_i = [\mathbf{E}_i]_{:, 1:S}^H \in \mathbb{C}^{S \times N}$. Unlike conventional digital architectures that require explicit matrix multiplications at each subcarrier, SIM physically realizes fully-analog beamforming in the wave domain by tuning the phase shifts and transmission coefficients of metasurfaces. 

Consequently, the resulting end-to-end channel matrix $\mathbf{H}_i\in \mathbb{C}^{S\times S}$ for the $i$-th subcarrier is expressed as: 
\begin{equation}
\mathbf{H}_i=\mathbf{Q}_i\,\mathbf{G}_i\,\mathbf{P}_i= [\mathbf{\Lambda}_i]_{1:S,1:S},
\end{equation}
where $[\mathbf{\Lambda}_i]_{1:S,1:S}\in \mathbb{C}^{S\times S}$ is a diagonal matrix capturing the $S$ dominant eigenmodes of $\mathbf{\Lambda}_i$. 

In conventional MIMO OFDM, imperfect precoding and combining lead to residual IAI due to the off-diagonal elements in $\mathbf{G}_i$, which distort the received signals across subcarriers~\cite{IAI}. However, the proposed SIM-enhanced architecture eliminates IAI by performing beamforming and combining directly in the EM wave domain. As illustrated in Fig.~\ref{SIM}, the SIM structure effectively forms independent parallel spatial subchannels in the analog EM wave domain, ensuring natural decoupling between data streams. 
Unlike traditional approaches that rely on baseband digital SVD to approximate a diagonalized channel, SIM dynamically reconfigures the propagation environment, inherently suppressing IAI by physically steering each stream through independent transmission paths.

By applying the fully-analog precoding $\mathbf{P}_i$ and combining $\mathbf{Q}_i$ within the proposed system, the end-to-end channel matrix $\mathbf{H}_i$ becomes a strictly diagonal form:
\begin{equation}
\mathbf{H}_i =
\begin{bmatrix}
[\lambda_{i}]_1 & 0 & \cdots & 0 \\
0 & [\lambda_{i}]_2 & \cdots & 0 \\
\vdots & \vdots & \ddots & \vdots \\
0 & 0 & \cdots & [\lambda_{i}]_S
\end{bmatrix}.
\end{equation}
 
Since $\mathbf{P}_i$ and $\mathbf{Q}_i$ act directly on each subcarrier, employing a frequency-domain model can simplify the representation of the system by avoiding explicit IFFT/FFT computations while preserving the underlying OFDM structure~\cite{fully-digital4}, \cite{Fchannel}. Thus, the received frequency-domain OFDM signal $\mathbf{y}_i\in \mathbb{C}^{S\times1}$ of the $i$-th subcarrier is expressed as:
\begin{equation} 
\mathbf{y}_i = \mathbf{H}_i\,\mathbf{x}_i + \mathbf{Q}_i\,\mathbf{N}_i=\mathbf{Q}_i\,\mathbf{G}_i\,\mathbf{P}_i\,\mathbf{x}_i + \mathbf{Q}_i\,\mathbf{n}_i,
\end{equation} 
where $\mathbf{x}_i\in \mathbb{C}^{S\times1}$ represents the OFDM frequency-domain symbol vector of the $i$-th subcarrier and $\mathbf{n}_i\in \mathbb{C}^{S\times1}$ is the Gaussian noise vector satisfying $\mathbf{n}_i\sim \mathcal{CN}(\mathbf{0},\sigma_n^2\,\mathbf{I}_S)$.

\begin{remark}
\textnormal{By performing precoding and combining directly in the EM wave domain, the proposed system achieves fully-analog spatial multiplexing within the SIM structure, eliminating the extra need for power-intensive baseband digital processing. Since IAI is eliminated at the physical level, each data stream can be transmitted and received independently without requiring complex baseband digital processing. This makes the SIM-enhanced architecture highly scalable and energy-efficient for wideband MIMO OFDM applications.}
\end{remark}

\section{Problem Formulation and Solution of the Proposed Wideband SIM for MIMO OFDM}
~~In this section, the optimization problem is formulated to address the challenges of multi-carrier transmission in frequency-selective channels. Following the problem formulation, a comprehensive solution framework is presented to refine the SIM parameters iteratively to achieve the desired fully-analog beamforming function in the wave domain. 
\label{Section3}
\subsection{Problem Formulation}
~~The multi-carrier optimization involves adjusting the phase shift matrices of the TX-SIM $\theta_m^l$ and RX-SIM $\zeta_n^k$ to ensure that the channel matrix $\mathbf{H}$ of the $N_c$ subcarriers closely approximates the targeted diagonal matrix $\mathbf{\Lambda}_{1:S,1:S}$ across the system bandwidth $B$. The fitting error for the desired channel alignment is characterized by the sum of the Frobenius norms subject to (s.t.) several constraints, leading to the following optimization problem:
\begin{subequations}
\label{OF_original}
\begin{align}
\label{Opt_original}
\mathcal{P}_1:\min_{\theta_m^l,~\zeta_n^k,~\alpha} & \mathrm{\Gamma}=\sum_{i=1}^{N_c}{||\alpha\mathbf{Q}_i\,\mathbf{G}_i\,\mathbf{P}_i-[\mathbf{\Lambda}_i]_{1:S,1:S}||}_F^2  \\
\text{s.t.} \quad &\mathbf{P}_i=\mathbf{\Phi}^L\mathbf{W}^L_i...\mathbf{\Phi}^2\mathbf{W}^2_i\mathbf{\Phi}^1\mathbf{W}^1_i,~i\in \mathcal{N}_c, \label{st.P} \\
     &\mathbf{Q}_i=\mathbf{U}^1_i\mathbf{\Psi}^1\mathbf{U}^2_i\mathbf{\Psi}^2...\mathbf{U}^K_i\mathbf{\Psi}^K,~i\in \mathcal{N}_c,  \label{st.Q}\\
     &\mathbf{\Phi}^l={\rm diag}([\phi_1^l,  \phi_2^l, \ldots, \phi_M^l]^T),~l\in \mathcal{L}, \label{st.pPhi} \\
     &\mathbf{\Psi}^k={\rm diag}([\psi_1^k,  \psi_2^k, \ldots, \psi_N^k]^T), k\in \mathcal{K}, \label{st.pPsi} \\
     &|\phi_m^l|=|e^{j\theta_m^l}|=1,~m\in \mathcal{M},~l\in \mathcal{L},  \label{st.phi}\\
     &|\psi_n^k|=|e^{j\zeta_n^k}|=1,~n\in \mathcal{N},~k\in \mathcal{K}, \label{st.psi}\\
     &\alpha\in \mathbb{C},  \label{st.alpha}
\end{align}
\end{subequations}
where $\alpha$ is a complex scaling factor compensated by SIM. 

\begin{remark}
\textnormal{The scaling factor $\alpha$ aims to compensate for the adaptive gain in the SIM-enhanced architecture. Again, unlike conventional architectures relying on baseband digital precoding and combining, the SIM-enhanced fully-analog beamforming system avoids additional hardware overhead. These energy-saving parts can be utilized to compensate for the energy loss caused by passing through the SIM.}
\end{remark}

\begin{table}[t]
\centering
\caption{\centering{\protect\\{\textsc{Optimization results when considering different setups of system bandwidth.}}}}\
\label{IFT} 
	\setlength{\tabcolsep}{0.8pt} 
	\renewcommand\arraystretch{1.3} 
\newcommand\xrowht[2][0]{\addstackgap[.5\dimexpr#2\relax]{\vphantom{#1}}}
\label{ofdmss}
\begin{tabular}{|m{1.5cm}<{\centering}|m{3.5cm}<{\centering}|m{3.5cm}<{\centering}|}
\hline
 & {\textbf{Interference-free transmission}} & \textbf{Interference-present transmission}  \\ \hline 
\textbf{Bandwidth} &$B\le B_e$  &$B> B_e$ \\ \hline 
\textbf{Fitting error} & $\mathrm{\Gamma}_{\min}\approx0$ & $\mathrm{\Gamma}_{\min}>0$\\ \hline 
\textbf{Illustration of the end-to-end channel} &\includegraphics[width=0.18\textwidth]{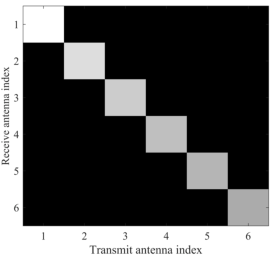} & \includegraphics[width=0.18\textwidth]{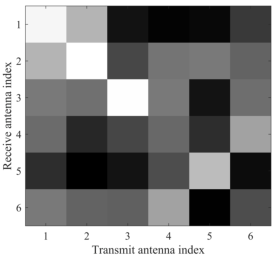}  \\ \hline 
\end{tabular}
\end{table}

However, direct optimization over the entire system bandwidth $B$ faces significant challenges due to the physical and EM constraints of SIM architectures. As illustrated in Table~\ref{IFT}, the effective bandwidth $B_e$ defines the range within which the SIM can achieve optimal diagonalization of the end-to-end channel matrix. When the system operates within $B\le B_e$, the fitting error $\mathrm{\Gamma}$ approaches zero, enabling interference-free parallel subchannels. However, where $B> B_e$, it may be hard to use SIMs to produce the desired diagonal channel response, and each data stream suffers from increased interference.

Therefore, to characterize the effective transmission bandwidth, we introduce an expected threshold $\varepsilon$ to ensure that the normalized fitting error $\Omega = \mathrm{\Gamma}/{||[\mathbf{\Lambda}_i]_{1:S,1:S}||}_F^2 \leq \varepsilon$ within $B_e$ for achieving IAI mitigation. Let $N_e$ denote the number of subcarriers that fall within the bandwidth $B_e$, with $f_i = f_0+(i - \frac{N_e+1}{2})\Delta f,~\Delta f = \frac{B_e} {N_e},~i\in \mathcal{N}_e$.

As a result, $\mathcal{P}1$ is reformulated as:
\begin{subequations}
\label{OF_P2}
\begin{align}
\mathcal{P}2:\max_{B_e,\theta_m^l,\zeta_n^k,~\alpha} & B_e, \label{Opt_Bo}\\
\text{s.t.} \quad
& \Omega =\frac{\sum_{i=1}^{N_e}{|| \alpha \mathbf{Q}_i \mathbf{G}_i \mathbf{P}_i - [\mathbf{\Lambda}_i]_{1:S,1:S} ||_F^2}}{{||[\mathbf{\Lambda}_i]_{1:S,1:S}||}_F^2} \leq \varepsilon, \label{st.Bo} \\
&\mathbf{P}_i=\mathbf{\Phi}^L\mathbf{W}^L_i...\mathbf{\Phi}^2\mathbf{W}^2_i\mathbf{\Phi}^1\mathbf{W}^1_i,~i\in \mathcal{N}_e, \label{st.Po} \\
     &\mathbf{Q}_i=\mathbf{U}^1_i\mathbf{\Psi}^1\mathbf{U}^2_i\mathbf{\Psi}^2...\mathbf{U}^K_i\mathbf{\Psi}^K,~i\in \mathcal{N}_e,  \label{st.Qo}\\
&\eqref{st.pPhi}-\eqref{st.alpha}, \label{st.allo}
\end{align}
\end{subequations}

In this paper, we adopt the bi-section method to identify the optimal effective bandwidth $B_e$ in which satisfactory interference cancellation is achieved. For a tentative $B_e$, Problem $\mathcal{P}_2$ is reduced to:
\begin{subequations}
\label{OF}
\begin{align}
\label{Opt}
\mathcal{P}_3:\min_{\theta_m^l,~\zeta_n^k,~\alpha} & \mathrm{\Gamma}=\sum_{i=1}^{N_e}{||\alpha\mathbf{Q}_i\mathbf{G}_i\mathbf{P}_i-[\mathbf{\Lambda}_i]_{1:S,1:S}||}_F^2  \\
\text{s.t.} \quad &\eqref{st.Po},~\eqref{st.Qo},~\eqref{st.allo}
\end{align}
\end{subequations}

\begin{remark} 
\textnormal{The reformulation from Problem $\mathcal{P}_1$ to $\mathcal{P}_3$ ensures that the SIM configuration achieves interference-free transmission across the maximum effective bandwidth.}
\end{remark}

\subsection{Solution for Phase Shift Optimization}
~~The Problem $\mathcal{P}_3$ in~\eqref{OF} is challenging to solve due to the highly coupled variables presented in the objective function $\mathrm{\Gamma}$ and the non-convex unit-modulus constraints imposed on each transmission coefficient. To address this challenge, we next transform the optimization objective function into a convex form and solve it by customizing a BCD-PCCP algorithm.
\vspace{-0.2 cm}
\begin{lemma}
\textnormal{By leveraging the matrix vectorization method, the objective function $\mathrm{\Gamma}$ can be transformed into a standard quadratic form w.r.t. w.r.t. $\bm{\phi}^l$ and $\bm{\psi}^k$, respectively, as shown in \eqref{st.ppp} and \eqref{st.qqq} at the top of the next page, where $\mathbf{P}_i = \mathbf{P}_i^\mathbb{L} \mathbf{\Phi}^l \mathbf{P}_i^\mathbb{R}$, $\mathbf{P}^\mathbb{L}_i=\mathbf{\Phi}^L\mathbf{W}^L...\mathbf{\Phi}^{l+1}\mathbf{W}^{l+1}\in \mathbb{C}^{M\times M}$, and $\mathbf{P}^\mathbb{R}_i=\mathbf{W}^l\mathbf{\Phi}^{l-1}\mathbf{W}^{l-1}...\mathbf{\Phi}^{1}\mathbf{W}^{1}\in \mathbb{C}^{M\times S}$. $\mathbf{Q}_i = \mathbf{Q}_i^\mathbb{L} \mathbf{\Psi}^k \mathbf{Q}_i^\mathbb{R}$, where $\mathbf{Q}^\mathbb{L}_i=\mathbf{U}^1\mathbf{\Psi}^1...$ $\mathbf{U}^{k-1}\mathbf{\Psi}^{k-1}\mathbf{U}^k\in \mathbb{C}^{S\times N}$ and $\mathbf{Q}^\mathbb{R}_i=\mathbf{U}^{k+1}\mathbf{\Psi}^{k+1}...\mathbf{U}^{K}\mathbf{\Psi}^{K}\in \mathbb{C}^{N\times N}$. $\mathbf{\Phi}^l = {\rm diag}(\bm{\phi}^l)$, where $\bm{\phi}^l = [\phi_1^l,  \phi_2^l, \ldots, \phi_M^l]^T, ~l\in \mathcal{L}$. $\mathbf{\Psi}^k = {\rm diag}(\bm{\psi}^k)$, where $\bm{\psi}^k = [\psi_1^k,  \psi_2^k, \ldots, \psi_N^k]^T,~k\in \mathcal{K}$.}
\end{lemma}

\begin{figure*}[!t]
\centering
\label{OF_new}
\begin{align}
\mathrm{\Gamma}=&\sum_{i=1}^{N_c}[\alpha^2({\bm{\phi}^l})^H({\mathbf{P}^\mathbb{L}_i}^H{\mathbf{G}_i}^H{\mathbf{Q}_i}^H\mathbf{Q}_i\mathbf{G}_i\mathbf{P}^\mathbb{L}_i)\odot(\mathbf{P}^\mathbb{R}_i{\mathbf{P}^\mathbb{R}_i}^H) {\bm{\phi}^l} -\alpha( {\rm vec}({\rm diag}(\bm{\phi}^l))^H)^H {\rm vec}(\mathbf{P}^\mathbb{R}_i[\mathbf{\Lambda}_i]_{1:S,1:S}^H\mathbf{Q}_i\mathbf{G}_i\mathbf{P}^\mathbb{L}_i) -  \nonumber\\ 
&\alpha^* ({\rm vec}({\rm diag}(\bm{\phi}^l)))^H{\rm vec}({\mathbf{P}^\mathbb{L}_i}^H\mathbf{G}_i^H\mathbf{Q}_i^H[\mathbf{\Lambda}_i]_{1:S,1:S}{\mathbf{P}^\mathbb{R}_i}^H)-{\rm Tr}([\mathbf{\Lambda}_i]_{1:S,1:S}[\mathbf{\Lambda}_i]_{1:S,1:S}^H)] \label{st.ppp} \\ 
=&\sum_{i=1}^{N_c}[\alpha^2({\bm{\psi}^k})^H({\mathbf{Q}^\mathbb{L}_i}^H\mathbf{Q}^\mathbb{L}_i) \odot (\mathbf{Q}^\mathbb{R}_i\mathbf{G}_i\mathbf{P}_i{\mathbf{P}_i}^H{\mathbf{G}_i}^H{\mathbf{Q}^\mathbb{R}_i}^H){\bm{\psi}^k}-\alpha ({\rm vec}({\rm diag}(\bm{\psi}^k))^H)^H {\rm vec}(\mathbf{Q}^\mathbb{R}_i\mathbf{G}_i\mathbf{P}_i[\mathbf{\Lambda}_i]_{1:S,1:S}^H\mathbf{Q}^\mathbb{L}_i) - \nonumber\\ 
&\alpha^* ({\rm vec}({\rm diag}({\bm{\psi}^k})))^H{\rm vec}({\mathbf{Q}^\mathbb{L}_i}^H[\mathbf{\Lambda}_i]_{1:S,1:S}\mathbf{P}_i^H\mathbf{G}_i^H{\mathbf{Q}^\mathbb{R}_i}^H)-{\rm Tr}([\mathbf{\Lambda}_i]_{1:S,1:S}[\mathbf{\Lambda}_i]_{1:S,1:S}^H)]. \label{st.qqq} 
\end{align}
\hrule 
\end{figure*}
\begin{proof}
\textnormal{Please see Appendix A.}
\renewcommand{\qedsymbol}{}
\end{proof}

Since $\mathrm{\Gamma}$ in \eqref{st.ppp} is a convex second-order cone programming (SOCP) problem w.r.t. ${\bm{\phi}^l}$, the optimization problem can be efficiently solved to obtain the optimal phase shifts for the $l$-th layer of the TX-SIM. The overall optimization is conducted iteratively in a layer-by-layer manner across all $L$ layers of the TX-SIM, with a similar approach applied to optimize the $K$ layers of the RX-SIM. However, since the given constraints \eqref{st.phi} and \eqref{st.psi} contain a non-convex constraint, the traditional convex optimization method cannot be directly used. The PCCP method is exploited at each iteration to relax the non-convex constraints with convex surrogates by leveraging the PCCP method~\cite{CCP}. The BCD framework refines the values of the phase shifts $\bm{\phi}^l$ and $\bm{\psi}^k$ in alternating cycles, keeping one fixed while optimizing the other at each iteration~\cite{BCD}.

To address the non-convexity of the unit-modulus constraints of the TX-SIM's $l$-th layer $|\phi_m^l|=|e^{j\theta_m^l}|=1$, these constraints are equivalently reformulated as ${1 \leq |\phi_m^l|}^2 \leq 1,~m\in \mathcal{M}$~\cite{PCCP}. By introducing the non-negative slack variable $\bm{\imath}=[\imath_1, \imath_2, \ldots, \imath_{2M}]^T\in {\mathbb{R}_{\geq 0}}^{2M\times 1}$, the unit-modulus constraint \eqref{st.phi} of the TX-SIM can be tackled by PCCP:
\begin{align}
{|\phi_m^l|}^2\le1+\imath_{M+m},~m\in \mathcal{M}, \label{CCP_P1} \\
{|\phi_0^l|}^2-2\Re{({\phi_m^l}^*{\phi_0^l})}\le\imath_{m}-1,~m\in \mathcal{M}, \label{CCP_P2}
\end{align}
where $\phi_0^l$ is the reference phase at the $l$-th layer of TX-SIM.

Similarly, another set of non-negative slack variables, $\bm{\jmath} = [\jmath_1, \jmath_2, \ldots, \jmath_{2N}]^T \in {\mathbb{R}_{\geq 0}}^{2N\times 1}$, is introduced to handle the $k$-th layer of RX-SIM's unit-modulus constraint $|\psi_n^k|=|e^{j\zeta_n^k}|=1$. The unit-modulus constraint \eqref{st.psi} is addressed as follows:
\begin{align}
{|\psi_n^k|}^2\le1+\jmath_{N+n},~n\in \mathcal{N}, \label{CCP_Q1} \\
{|\psi_0^k|}^2-2\Re{({\psi_n^k}^*{\psi_0^k})}\le\jmath_{n}-1,~n\in \mathcal{N}, \label{CCP_Q2}
\end{align}
where $\psi_0^k$ is the reference phase at the $k$-th layer of RX-SIM. 

However, adding $\bm{\imath}$ and $\bm{\jmath}$ to relax the unit-modulus constraint may cause these constraints to be violated. Therefore, the penalty factors $\rho$ and $\varrho$ are incorporated into the objective function $\mathrm{\Gamma}$ to minimize the values of the auxiliary variables.

Let ${\bm{\theta}^l} = \{\theta_m^l\}_1^M,~l\in \mathcal{L}$ and ${\bm{\zeta}^k} = \{\zeta_n^k\}_1^N,~k\in \mathcal{K}$. As a result, Problem $\mathcal{P}_3$ in \eqref{OF} is transformed into
\begin{subequations}
\label{OF_all}
\begin{align}
\mathcal{P}_4:
\min_{{\bm{\theta}^l},~{\bm{\zeta}^k},~\alpha,~\bm{\imath},~\bm{\jmath}}~&\mathbf{F} ({\bm{\theta}^l},{\bm{\zeta}^k}) = \mathrm{\Gamma} + \rho \sum_{i=1}^{2M}{\imath_{i}} + \varrho \sum_{i=1}^{2N}{\jmath_{i}}, \label{obj_function}\\
\text{s.t.} \quad
&{\bm{\phi}^l} = e^{j{\bm{\theta}^l}},~l\in \mathcal{L}, \label{st.ppPhi}  \\
&{\bm{\psi}^k} = e^{j{\bm{\zeta}^k}},~k\in \mathcal{K}, \label{st.ppPsi} \\
&\eqref{st.alpha},~\eqref{CCP_P1},~\eqref{CCP_P2},~\eqref{CCP_Q1},~\eqref{CCP_Q2}, \label{st.all} \\
& \imath_{i}\ge0,~i=1,~2, \ldots, 2M, \label{st.penaltyP}\\
& \jmath_{i}\ge0,~i=1,~2, \ldots, 2N, \label{st.penaltyQ} \\
& \rho,~\varrho \in \mathbb{R},   \label{st.rho}
\end{align}
\end{subequations}

Since the optimization problem \eqref{obj_function} involves multiple sets of variables, we solve it by using BCD to tackle one subblock while keeping others fixed at each iteration. 

\begin{enumerate}[Step 1]
\item  When other variables except $\{{\bm{\theta}^l},~\bm{\imath}\}$ are fixed, Problem $\mathcal{P}_4$ can be expressed into Problem $\mathcal{P}_5$ w.r.t. $\{{\bm{\theta}^l},~\bm{\imath}\}$:
\begin{subequations}
\label{OF_P}
\begin{align}
\mathcal{P}_5:  ~
\min_{{\bm{\theta}^l},~\bm{\imath}}\quad&\mathrm{\Gamma}({\bm{\theta}^l})=\mathrm{\Gamma}+\rho \sum_{i=1}^{2M}{\imath_{i}} \label{obj_P}\\
\text{s.t.} \quad
 &\eqref{st.ppPhi},~\eqref{st.all},~\eqref{st.penaltyP},~\eqref{st.rho}
\end{align}
\end{subequations}
Problem $\mathcal{P}_5$ is solved iteratively using the CVX toolbox, optimizing each metasurface layer independently. Specifically, at each iteration, the updated variables ${\bm{\theta}^l}$ containing phase shifts of $M$ meta-atoms at the $l$-th layer are updated while maintaining fixed configurations for all other layers.

\item Similarly, when other variables except $\{{\bm{\zeta}^k},\bm{\jmath}\}$ are fixed, Problem $\mathcal{P}_4$ simplifies to \eqref{st.qqq}:
\begin{subequations}
\label{OF_Q}
\begin{align}
\mathcal{P}_6:  ~
\min_{{\bm{\zeta}^k},~\bm{\jmath}}\quad&\mathrm{\Gamma}({\bm{\zeta}^k})=\mathrm{\Gamma}+\varrho \sum_{i=1}^{2N}{\jmath_{i}} \label{obj_Q}\\
\text{s.t.} \quad
 &\eqref{st.ppPsi},~\eqref{st.all},~\eqref{st.penaltyQ},~\eqref{st.rho}
\end{align}
\end{subequations}

The CVX toolbox can be readily employed here to optimize the phase shifts ${\bm{\zeta}^k}$ with $N$ meta-atoms at the $k$-th layer of the RX-SIM, iterating over each layer while fixing the configurations of all others.

\item At each iteration, the scaling factor $\alpha$ is also updated. Specifically, calculating the partial derivative of the objective function $\mathrm{\Gamma}$ w.r.t. $\alpha$ and setting the derivative $\partial {\mathrm{\Gamma}}/\partial {\alpha}$ to 0 yields:
\begin{align}
\alpha = \frac{\sum_{i=1}^{N_e}[\Re({\rm Tr}([\mathbf{\Lambda}_i]_{1:S,1:S}^H\,\mathbf{Q}_i\,\mathbf{G}_i\,\mathbf{P}_i))]}{{\sum_{i=1}^{N_e}||\mathbf{Q}_i\,\mathbf{G}_i\,\mathbf{P}_i||}_F^2}.
\label{update_a}
\end{align}
\end{enumerate}

The whole optimization of Problem $\mathcal{P}_4$ alternates between the subblocks of $\bm{\phi}^l$ and $\bm{\psi}^k$, ensuring convergence to a solution after $L$ and $K$ iterations for the TX-SIM and RX-SIM, respectively. Algorithm \ref{CCP} shows the overall process to solve Problem $\mathcal{P}_4$.

\begin{algorithm}
\caption{The Proposed BCD-PCCP Algorithm for Optimizing Phase Shifts for Analog Beamforming}
\label{CCP}
\textbf{Initialization:} Set the initial phase shift matrices ${\bm{\theta}^l}^{[0]}$ and ${\bm{\zeta}^k}^{[0]}$, scaling factor $\alpha$, the maximum penalty factors $\rho_{max}$ and $\varrho_{max}$, the maximum outer iteration number ${\tau}_{max}$, and the maximum inner iteration number ${\tau_\theta}_{max}$ and ${\tau_\zeta}_{max}$. \\
\Repeat {Convergence criteria are met}
{
\If{$\tau<{\tau}_{max}$}
{
     Initialize ratio coefficient $ \mu > 1 $, and set $\tau_\theta = \tau_\zeta = 0 $;\\
\For{ $i = 1, \dots, N_e$}
{
\For{ $l = 1, \dots, L$}
{ 
	\eIf{$\tau_\theta<{\tau_\theta}_{max}$}
		{
         Update ${\bm{\theta}^l}^{[\tau_\theta+1]}$ using the PCCP from adapted Problem $\mathcal{P}_5$ in \eqref{OF_P}:\\
         Update $\rho = \min\{\mu\rho, {\rho}_{\text{max}}\}$;\\
         Increment $ \tau_\theta = \tau_\theta + 1 $;\\
		}
		{
		 Reinitialize ${\bm{\theta}^l}^{[0]}$ randomly. Set up $\mu > 1$ and reset $\tau_\theta = 0$.
		}
}
}
\For{ $i = 1, \dots, N_e$}
{
\For{ $k = 1, \dots, K$}
{
	{
	\eIf{$\tau_\zeta<{\tau_\zeta}_{max}$}
		{
         Update ${\bm{\zeta}^k}^{[\tau_\zeta+1]}$ using the PCCP from adapted Problem $\mathcal{P}_6$ in \eqref{OF_Q}:\\
         Update $\varrho = \min\{\mu\varrho, {\varrho}_{\text{max}}\}$;\\
         Increment $ \tau_\zeta = \tau_\zeta + 1 $;\\
		}
		{
            Reinitialize ${\bm{\zeta}^k}^{[0]}$ randomly. Set up $\mu > 1$ and reset $\tau_\zeta = 0$.
		}
    }
}
}
Increment $\tau=\tau+1$; \\
Update transmission coefficients ${\mathbf{P}_i}$ and ${\mathbf{Q}_i}$; \\
Update scaling factor $\alpha$ via \eqref{update_a};
}
}
Output ${\bm{\theta}^l} = {\bm{\theta}^l}^{[\tau_\theta+1]}$ and ${\bm{\zeta}^k} = {\bm{\zeta}^k}^{[\tau_\zeta+1]}$.
\end{algorithm}

\begin{remark}
\textnormal{The penalty factors $\rho$ and $\varrho$ are dynamically adjusted to balance objective minimization with constraint feasibility. To speed up convergence, these factors are increased by a ratio coefficient $\mu > 1$ during inner iterations, with appropriate upper bounds to avoid numerical issues.}
\end{remark}

\section{Performance Analysis of the Proposed BCD-PCCD Algorithm}
\label{Section4}
~~This section provides a comprehensive performance evaluation of Algorithm \ref{CCP} introduced in Section~\ref{Section3}, including convergence analysis and computational complexity analysis.
\subsection{Convergence Analysis}
\label{Section4.A}
~~Considering the convexity of \eqref{st.ppp} w.r.t. ${\bm{\theta}^l}$ for fixed ${\bm{\zeta}^k}$, the convergence properties of BCD-PCCP guarantee a monotonic decrease in the objective function at each iteration, and vice versa w.r.t. ${\bm{\zeta}^k}$. Thus, the proposed BCD-PCCP method reduces $\mathbf{F} ({\bm{\theta}^l},{\bm{\zeta}^k})$ by iteratively refining the convex approximations for ${\bm{\theta}^l}$ while keeping ${\bm{\zeta}^k}$ fixed. For each optimization step (either ${\bm{\theta}^l}$ or ${\bm{\zeta}^k}$), the convergence criteria are defined in a unified form.

\begin{align}
\label{tolerance_combined}
\left| \frac{\mathrm{\Gamma} (\bm{x}^{[\tau+1]}) - \mathrm{\Gamma} (\bm{x}^{[\tau]})}{\mathrm{\Gamma} (\bm{x}^{[\tau]})} \right| \leq \epsilon_1, \\
| \bm{x}^{[\tau+1]} - \bm{x}^{[\tau]} |_2 \leq \epsilon_2, \\
|| \bm{s} ||_1 \leq \epsilon_3,
\end{align}
where ${\bm{x}}$ represents either ${\bm{\theta}^l}$ or ${\bm{\zeta}^k}$, and ${\bm{s}}$ is the corresponding slack variables ${\bm{\imath}}$ or ${\bm{\jmath}}$. Parameters ${\epsilon_1}$, ${\epsilon_2}$, and ${\epsilon_3}$ are positive tolerance factors controlling the stability of the solution. $||\bm{s}||_1$ is the penalty term in either sub-objective function. 

Since $\rho\sum_{i=1}^{2M}{\imath_{i}}\approx~0$ and $\varrho\sum_{i=1}^{2N}{\jmath_{i}}\approx0$ as PCCP iterations progress~\cite{CCP}, each block update phase shifts towards an optimal solution for the partially minimized objective function. The convergence of the overall process is achieved when there is no significant improvement in $\mathbf{F} ({\bm{\theta}^l},{\bm{\zeta}^k})$ or the changes in the variable blocks are below a preset threshold.,
\begin{align}
\left |\frac{\mathbf{F} _{}({\bm{\theta}^l}^{[\tau+1]}, {\bm{\zeta}^k}^{[\tau+1]})-\mathbf{F} _{}({\bm{\theta}^l}^{[\tau]}, {\bm{\zeta}^k}^{[\tau]})}{\mathbf{F} _{}({\bm{\theta}^l}^{[\tau]}, {\bm{\zeta}^k}^{[\tau]})}\right | \leq \nu,
\end{align}
where $\nu > 0$ is a predefined threshold. 

Each step in the iteration process, first updating $\bm{\theta}^l$, then updating $\bm{\zeta}^k$, results in a monotonic decrease or stability in $\mathbf{F} ({\bm{\theta}^l},{\bm{\zeta}^k})$ to guarantee convergence. If convergence is not achieved, the algorithm is restarted with a new random phase matrix initialization to solve Problems $\mathcal{P}_5$ and $\mathcal{P}_6$ again.
\subsection{Complexity Analysis}
~~This part provides a detailed analysis of the computational complexity of the BCD-PCCP algorithm. The computational burden of Algorithm~\ref{CCP} mainly arises from solving the SOCP Problem $\mathcal{P}_4$ in~\eqref{OF_all}.

The major computational complexity using the PCCP method stems from solving the SOCP Problems $\mathcal{P}_5$ in~\eqref{OF_P} and $\mathcal{P}_6$ in \eqref{OF_Q}. Based on \cite{SOC1}, the complexity of an SOCP problem is $O(XY^{3.5} + X^{3.5}Y^{2.5})$, where $X$ represents the number of second-order cone (SOC) constraints and $Y$ represents the corresponding dimension. In solving the SOCP problem \eqref{OF_P}, there are $2M$ SOC constraints with dimension one for each inner PCCP iteration, where all $M$ meta-atoms in the $L$ layer of the TX-SIM must be updated iteratively. Similarly, there are $2N$ SOC constraints with dimension one in the SOCP problem \eqref{OF_Q}, which need to be iteratively updated across $K$ layers in the RX-SIM. Thus, the complexity in solving the SOCP problems \eqref{OF_P} and \eqref{OF_Q} in each subcarrier is $O(LM\times(2M)^{3.5})$ and $O(KN\times(2N)^{3.5})$, respectively.

By omitting lower-order terms, the overall computational complexity per iteration of solving Problem $\mathcal{P}_4$ in~\eqref{OF_all} is dominated by the SOCP complexity. Hence, the maximum computational complexity order of the Algorithm~\ref{CCP} per iteration is $O(N_e\times\tau_{max}\times[{\tau_\theta}_{max}\times(LM\times(2M)^{3.5})+{\tau_\zeta}_{max}\times(KN\times(2N)^{3.5})])$.

\section{Results and Analysis}
\label{Section5}
~~The numerical results are conducted in this section to characterize the performance of the proposed SIM-enhanced MIMO OFDM communication system.
\subsection{Simulation Configurations and Evaluation Criteria}
~~The simulation setup consists of metasurface layers for both TX-SIM and RX-SIM, each with a thickness of $D_T=D_R=0.05$ m. The spacing between the layers of TX-SIM and RX-SIM is set as $d_T = D_T/L$ and $d_R = D_R/K$, respectively. A center frequency $f_0=28$ GHz is chosen for simulations. The initial parameters of the SIM-enhanced system are configured with $S = 4$, $M = N = 100$, $L = K = 7$, $r_{T}=r_{R}= c/(2f_0)$, and $S_{T}=S_{R}= c^2/{(2f_0)}^2$. The TX-RX distance is set at 250 m with 100 scatterers randomly generated to model the multi-path channel characteristics. The simulation assumes an antenna gain of $9$ dBi, a system loss of $3$ dB, a total transmit power $P_t=20$ dBm, and a receiver noise sensitivity of $\sigma_n^2=-110$ dBm.

For the simulation of the proposed algorithm, 100 random initializations are performed to ensure robustness. For each initialization, the number of maximum outer iterations is set to ${\tau}_{max}=50$, and the inner optimization iterations ${\tau_\theta}_{max}=30$ and ${\tau_\zeta}_{max}=30$, respectively. To terminate the algorithm, the tolerance thresholds are set as $\epsilon_1 = 10^{-3}$,  $\epsilon_2 = 10^{-5}$, $\epsilon_3 = 10^{-5}$, and $\nu=10^{-6}$. The Monte Carlo method is utilized to average the results over 100 simulation runs.

Performance is assessed primarily through the normalized mean square error (NMSE) between the end-to-end channel matrices and target diagonal matrices:
\begin{align}
\mathcal{S}=\mathbb{E} (\sum_{i=1}^{N_e}\frac{{||\alpha\mathbf{Q}_i\,\mathbf{G}_i\,\mathbf{P}_i-[\mathbf{\Lambda}_i]_{1:S,1:S}||}_F^2}{{||[\mathbf{\Lambda}_i]_{1:S,1:S}||}_F^2} ).
\end{align}

The channel capacity of the proposed SIM-enhanced MIMO OFDM system is also evaluated by considering joint power allocation in both frequency and spatial domains. The power allocated to the $s$-th spatial stream on the $i$-th subcarrier. denoted as $[p_i]_s$, is optimized using the water-filling algorithm and can be expressed as:
\begin{align}
[p_i]_s = \max(0,~\tau_p-\frac{\sigma^2}{[{\lambda}_i]_s}),
\end{align}
where $\tau_p$ is the water-filling threshold, which ensures the total transmit power constraint is satisfied ${\textstyle \sum_{1}^{N_e}}  {\textstyle \sum_{1}^{S}} [p_i]_s = P_t$.

The spectral efficiency $\eta$ measures the efficiency of data transmission per unit bandwidth. Specifically, for the $i$-th subcarrier, the spectral efficiency $\eta_i$ can be calculated as:
\begin{align}
\eta_i = \sum^{S}_{s=1} \log_{2}{(1+\frac{[p_i]_s\,{|\alpha[{\lambda}_i]_s|}^2}{ {\textstyle \sum_{\tilde{s}\neq s }^{S}}[p_i]_{\tilde{s}}\,{|\alpha[h_{i}]_{s,{\tilde{s}}}|}^2+\sigma^2} )},
\end{align}
where $[h_{i}]_{s,{\tilde{s}}}$ represents the entry on the $s$-th row and the $\tilde s$-th column of $\mathbf{H}_i$, characterizing the inter-stream interference caused by the $\tilde{s}$-th stream on the $s$-th stream for the $i$-th subcarrier. Unlike conventional MIMO OFDM designs with extra digital combining, the proposed SIM-enhanced approach treats the residual signals from other data streams as interference.

The channel capacity $\mathcal{C}$ of the proposed SIM over the system bandwidth $B$ with $N_e$ subcarriers is expressed by: 
\begin{align}
\mathcal{C} = \sum_{i=1}^{N_e} \Delta f \cdot \eta_i.
\end{align}

\subsection{Performance Evaluation across OFDM Parameters}
\begin{figure}
	\centerline{\includegraphics[width=0.4\textwidth]{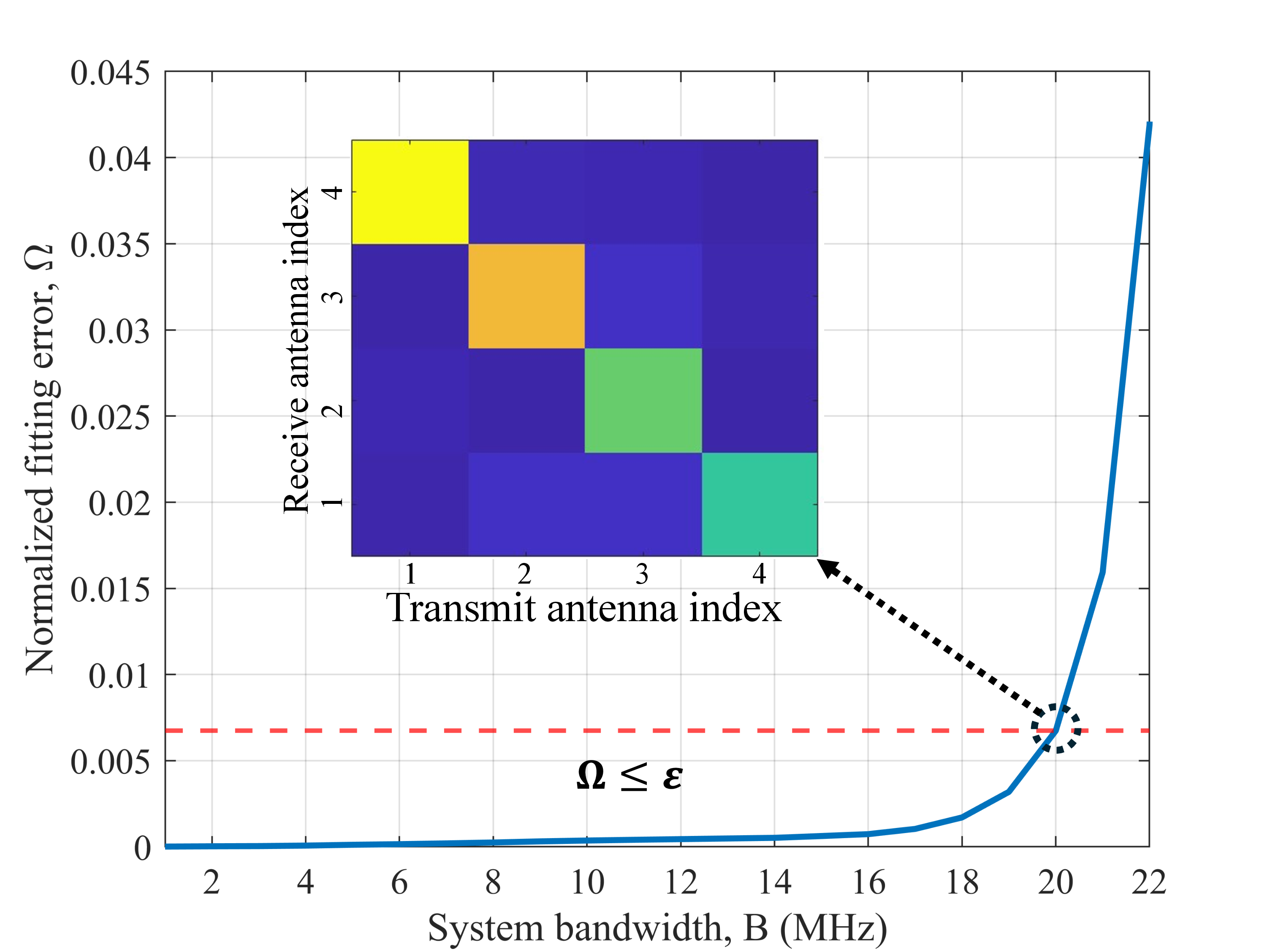}}
	\caption{{{Normalized fitting error $\Omega$ versus system bandwidth $B$, showing the maximum effective bandwidth $B_e$ where interference-free diagonalization is achieved.}}}
	\label{epsilon} 
\end{figure}

~~Fig.~\ref{epsilon} examines the maximum effective bandwidth $B_e$ of the proposed SIM-enhanced system, which defines the largest bandwidth over which interference-free diagonalization of the end-to-end channel is achieved. The horizontal axis shows the system bandwidth $B$, while the vertical axis indicates the normalized fitting error $\Omega$, which is computed for each $B$ by solving the constraint \eqref{st.Bo}. For $B<20$ MHz, $\Omega$ is consistently below the preset threshold $\varepsilon=0.0065$, indicating effective diagonalization of the channel matrix. This region demonstrates the system's robustness in achieving interference-free transmission. At $B=20$ MHz, $\Omega$ approaches $\varepsilon$, signifying the upper limit of the system’s ability to ensure interference-free diagonalization. To illustrate, we also visualize the end-to-end channel matrix at $B = 20$ MHz, showing that SIM achieves near-perfect diagonalization. However, as $B$ exceeds 20 MHz, the fitting error rises sharply, preventing the channel from maintaining the desired diagonalization due to inherent system constraints. This analysis highlights $B_e=20$ MHz as a key parameter for characterizing and optimizing the SIM-enhanced system. Within this effective bandwidth, the system achieves high spectral efficiency and channel capacity, while performance is significantly degraded for out-of-band signals. Identification of $B_e$ provides a precise metric to achieve interference-free transmission of the system to mitigate IAI.

\begin{figure}
	\centerline{\includegraphics[width=0.4\textwidth]{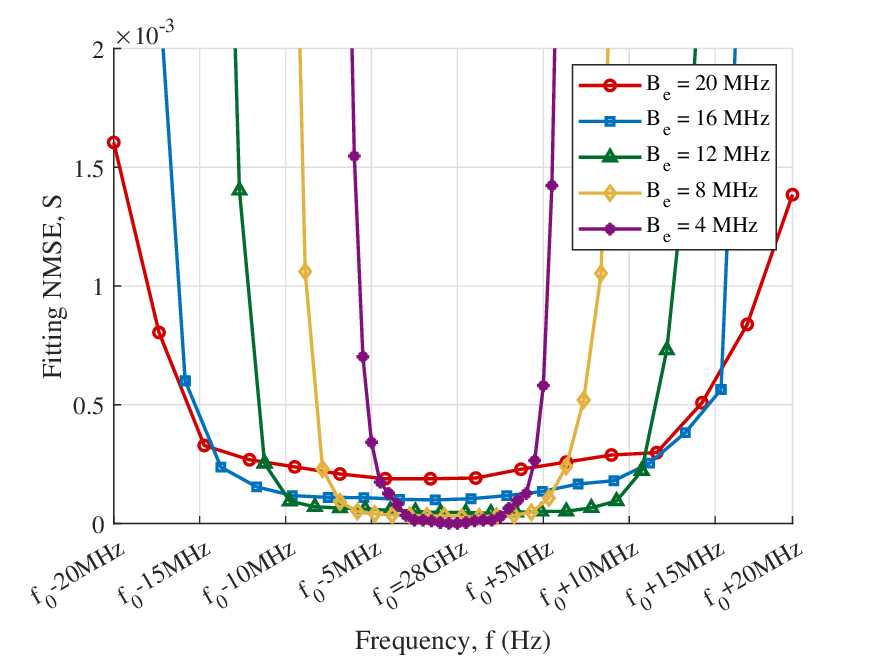}}
	\caption{{The per-subcarrier fitting NMSE $\mathcal{S}$ under different $B_e$ setups, where we consider a fixed subcarrier number of $N_e=16$ and varying subcarrier spacing.}}
	\label{changeBS} 
\end{figure}

Then, the fitting performance of the proposed SIM-enhanced system is explored under different subcarrier indices, where the subcarrier index value is related to $B_e$. The experimental results, depicted in Fig.~\ref{changeBS}, investigate the effect of varying subcarrier spacing $\Delta f$ on fitting NMSE $\mathcal{S}$ under a fixed number of subcarriers $N_c=16$. The study considers different configurations of $\Delta f=$1.25~MHz, 1 MHz, 750 kHz, 500 kHz, and 250 kHz, corresponding to an effective bandwidth of $B_e=\Delta f\cdot N_e=$20, 16, 12, 8, and 4 MHz, respectively. Fig.~\ref{changeBS} shows the fitting NMSE per-subcarrier within the designated bandwidth $B_e$, demonstrating that the proposed SIM-enhanced system effectively diagonalizes the channel matrix. The results show that for each $B_e$, the system maintains low and stable NMSE values, confirming the SIM's ability to mitigate inter-strean interference within their designated $B_e$. However, when $B_e>20$ MHz, fitting errors increase significantly and exceed the expected threshold, as demonstrated the same in Fig.~\ref{epsilon}, confirming that $B_e=20$ MHz is the maximum effective bandwidth achievable by the SIM setup considered.

\begin{figure}
	\centerline{\includegraphics[width=0.5\textwidth]{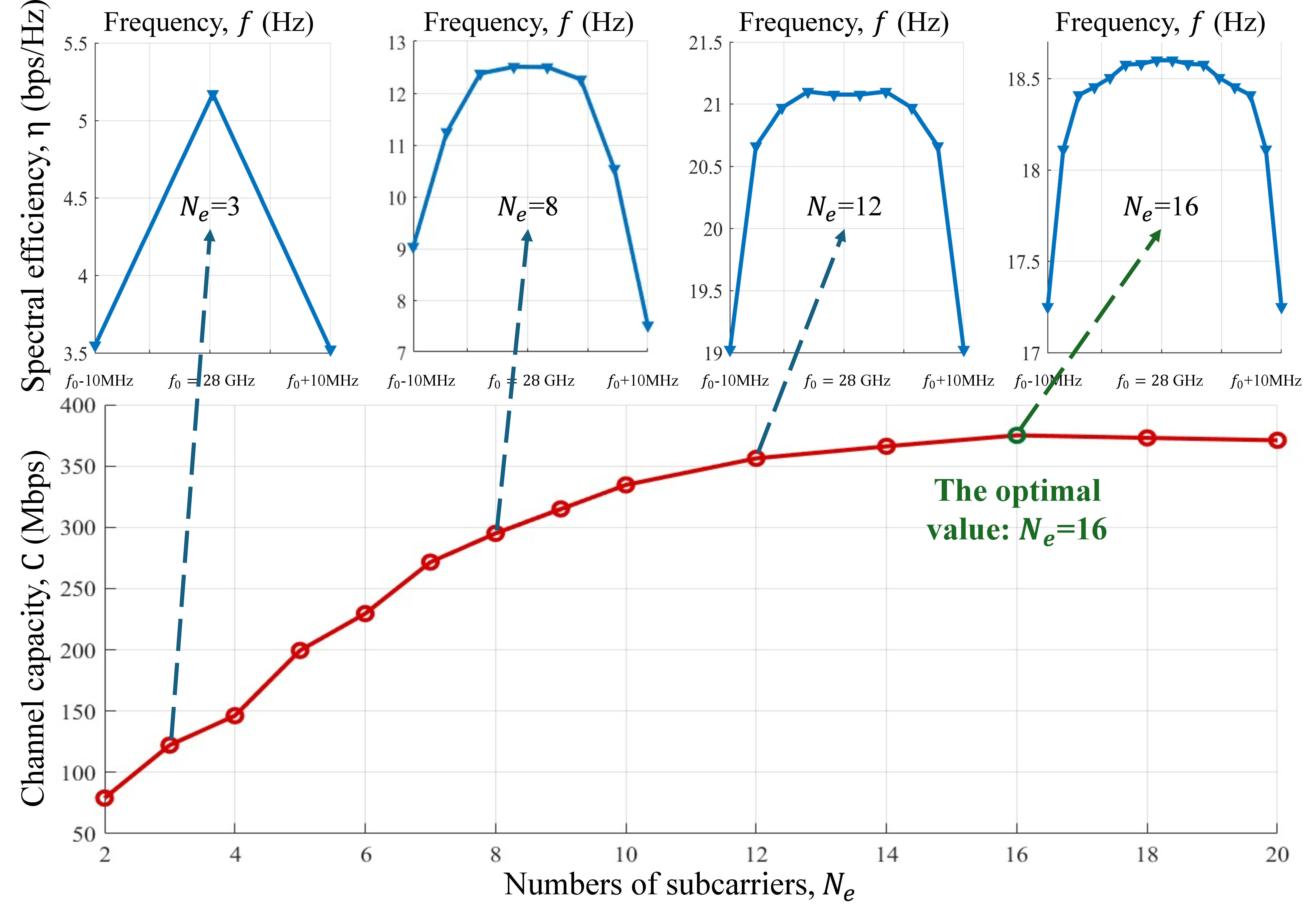}}
	\caption{{{The channel capacity $C$ versus the number of subcarriers $N_e$ at the fixed optimization bandwidth $B_e = 20$ MHz.}}}
	\label{changeNC} 
\end{figure}

Fig.~\ref{changeNC} illustrates the impact of varying the number of subcarriers on the channel capacity under a fixed optimization bandwidth $B_e = 20$ MHz. The bottom panel shows the channel capacity by calculating the sum of spectral efficiencies across the 20 MHz effective bandwidth, as shown in the top panel. The results demonstrate a non-linear trend: when $N_e<10$, the channel capacity improves significantly; at $N_e=16$, the capacity peaks at approximately 375 Mbps; when $N_e>16$, the capacity plateaus, with minimal improvement due to diminishing returns in spectral efficiency. This is because when $N_e$ is small, the large spacing $\Delta f$ makes it difficult for the SIM configuration to achieve effective fitting across the frequency spectrum, leading to suboptimal diagonalization and reduced spectral efficiency. In contrast, increasing the number of subcarriers improves frequency selectivity and diagonalization precision. However, beyond 16, the computational complexity of optimizing additional subcarriers outweighs the marginal capacity gains. These results show $N_e=16$ as the optimal configuration, striking a balance between computational complexity and system performance to maximize channel capacity and spectral efficiency within the effective bandwidth.

\subsection{Performance under Different System Parameters}
\begin{figure}
	\centerline{\includegraphics[width=0.4\textwidth]{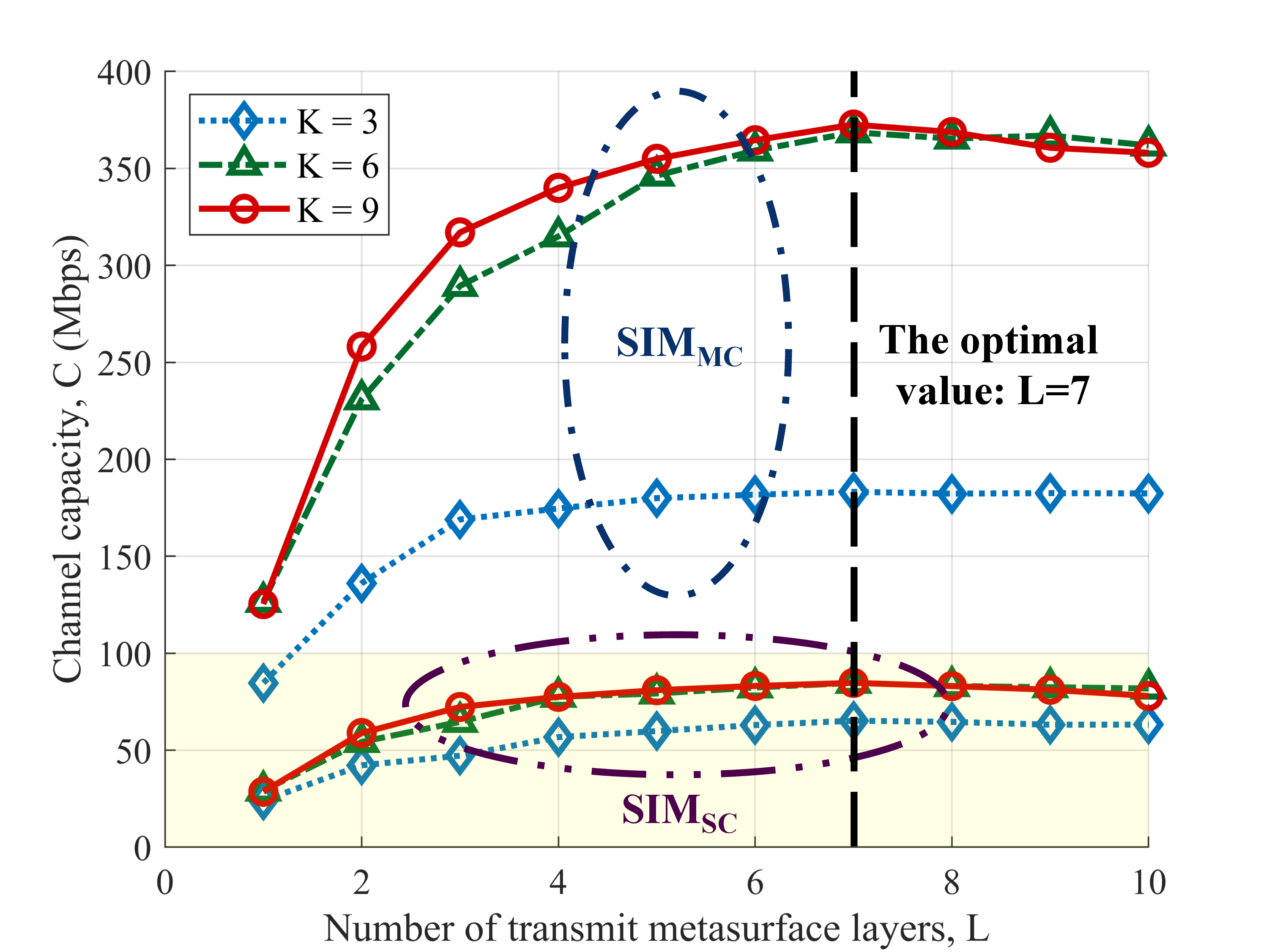}}
	\caption{{The channel capacity $C$ within effective bandwidth $B_e$ versus the number of transmit metasurface layers $L$ of SIM$_\text{MC}$ and SIM$_\text{SC}$ configurations under the setup of $M=N=100$.}}
	\label{changeLK} 
\end{figure}

~~Fig.~\ref{changeLK} evaluates the channel capacity versus the number of metasurface layers, where we consider two different SIM configuration methods: i) the proposed method considering multiple subcarriers (SIM$_\text{MC}$) and ii) the configuration only accounting for the single-carrier (SIM$_\text{SC}$) \cite{SIM-HMIMO}. The objective is to determine the optimal configuration of metasurface layers for achieving a balance between channel capacity and system complexity within the effective bandwidth $B_e=20$ MHz. The channel capacity of the baseline ii) is calculated by applying the SIM$_\text{SC}$ to signals outside the center frequency $f_0$ across $B_e$. Moreover, the three curves correspond to varying numbers of receive metasurface layers $K=3$, 6, and 9, demonstrating the influence of RX-SIM layers on system performance. The results show that SIM$_\text{MC}$ consistently outperforms SIM$_\text{SC}$. Specifically, SIM$_\text{SC}$ exhibits significantly lower channel capacity and marginal gains as $L$ increases. For SIM$_\text{MC}$, the channel capacity rises sharply with $L$, reaching its peak at $L=7$, beyond which additional layers result in diminishing returns. This is because excessive layer stacking reduces the ability of individual layers to effectively manipulate EM waves, as closer proximity yields a more diagonal inter-layer propagation matrix. The optimal configuration, $L=K=7$, balances channel capacity and system complexity, leveraging SIM$_\text{MC}$'s enhanced ability to mitigate multi-path effects and optimize wideband performance.

\begin{figure}
	\centerline{\includegraphics[width=0.4\textwidth]{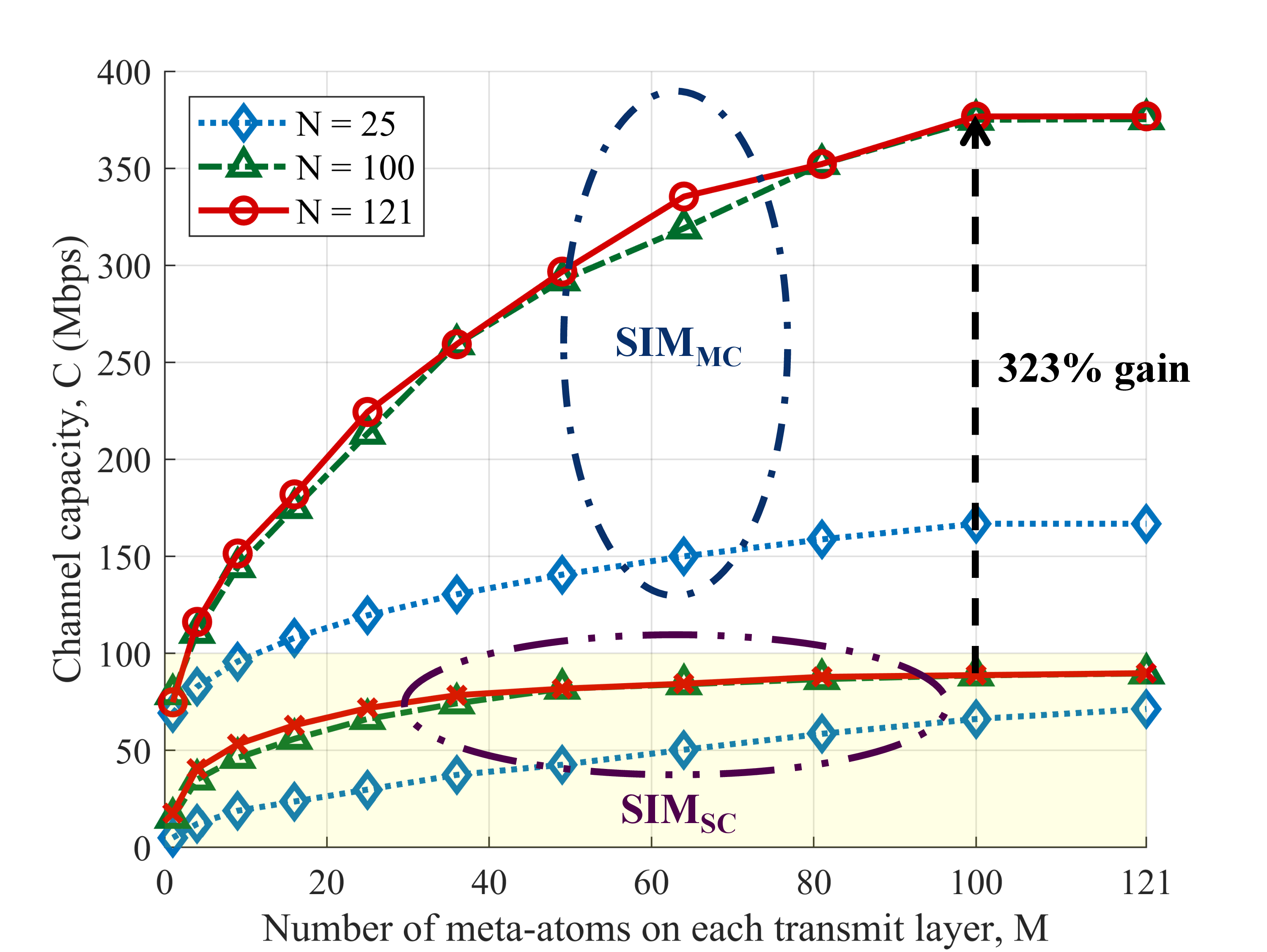}}
	\caption{{The channel capacity $C$ within effective bandwidth $B_e$ versus the number of meta-atoms $M$ on each transmit metasurface layer of two SIM configurations under the setup of $L=K=7$.}}
	\label{changeMN} 
\end{figure}

Next, the impact of the number of meta-atoms per layer is analyzed under the setup of $L=K=7$. Fig.~\ref{changeMN} demonstrates that the channel capacity increases with the number of meta-atoms per layer, as higher numbers enable more precise control of phase and amplitude adjustments across the wideband spectrum. However, this trend plateaus when the number of meta-atoms reaches $M=N=100$. Beyond this threshold, further increase in the meta-atom density yields negligible gains in channel capacity. This limitation arises from the physical properties of EM materials, optimization complexity, and intra-layer interactions, which restrict additional performance improvements. The optimal configuration, achieved with $M=N=100$ meta-atoms per layer and $L=K=7$ layers, offers a balanced trade-off between optimization complexity and performance. This configuration delivers a 323\% improvement in channel capacity over the SIM$_\text{SC}$ configuration within the effective bandwidth, highlighting the significant advantages of the proposed SIM$_\text{MC}$ in OFDM signal transmission. The results underscore the system's capability to efficiently optimize wideband communication performance while maintaining practical complexity.

\subsection{Validation of the Proposed Algorithm}
\begin{figure}
	\centerline{\includegraphics[width=0.45\textwidth]{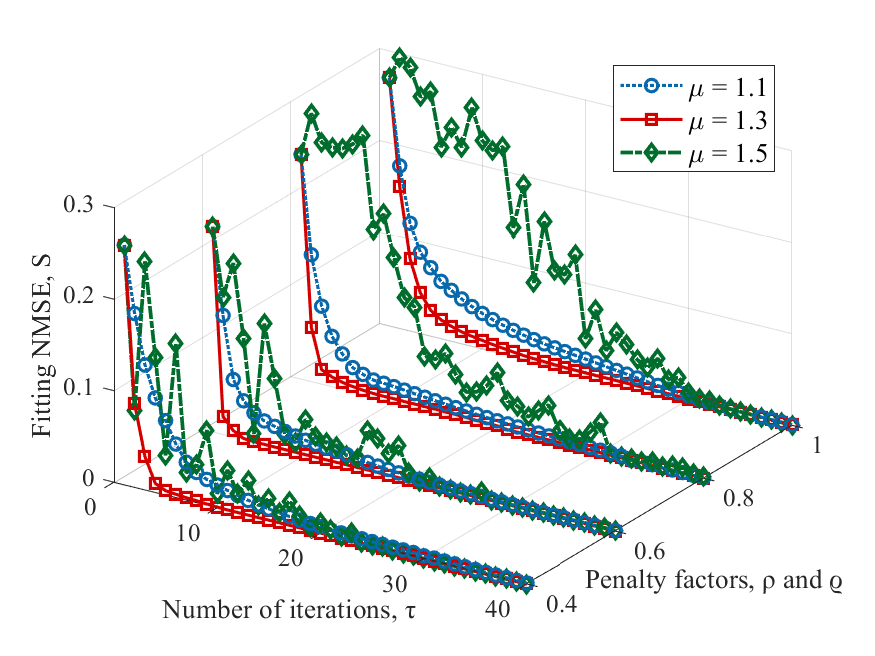}}
	\caption{{The convergence behavior of the fitting NMSE $\mathcal{S}$ versus the number of iterations $\tau$ under different penalty factors $\rho$ and $\varrho$.}}
	\label{iteration} 
\end{figure}

~~The fitting NMSE performance of the proposed BCD-PCCP algorithm is evaluated under various penalty factors $\rho$ and $\varrho$, as shown in Fig.~\ref{iteration}. The figure illustrates the convergence behavior of the fitting NMSE between the actual channel matrices and the target diagonal matrices within $B_e$, where the x-axis represents the number of iterations $\tau$ and the y-axis varies values of penalty factors $\rho$ and $\varrho$. Additionally, under the same setups of penalty factor, the performance is compared across different ratio coefficients $\mu$. The results indicate that the proposed algorithm consistently converges to a near-optimal solution under various setups of penalty factors and coefficients, demonstrating its robust convergence properties as established in Section~\ref{Section4.A}. However, excessively small or large penalty factors adversely impact the convergence speed, requiring more iterations to reach convergence. Furthermore, the convergence curve is smoother when $\mu=$1.1, compared to $\mu=$1.3, for fixed penalty factors. As iterations continue, increasing $\mu$ to 1.5 results in frequent fluctuations in the convergence curve due to overshooting. The optimal convergence performance is achieved with $\rho = \varrho = 0.6$ and $\mu = 1.3$. Nevertheless, the fitting NMSE of the proposed algorithm ultimately converges to the desired accuracy after sufficient iterations in all scenarios.

\begin{figure}[t]
    \centering
    \subfigure[SIM$_\text{SC}$\label{fig:subfig1}]{\includegraphics[width=0.25\textwidth]{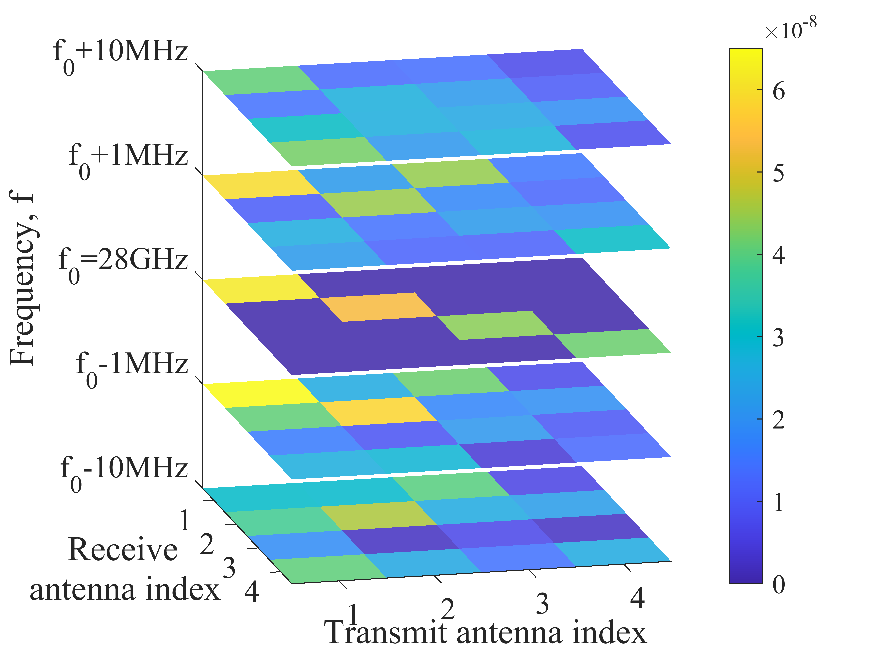}}
    \hspace{-0.5 cm}
    \subfigure[SIM$_\text{MC}$\label{fig:subfig2}]{\includegraphics[width=0.25\textwidth]{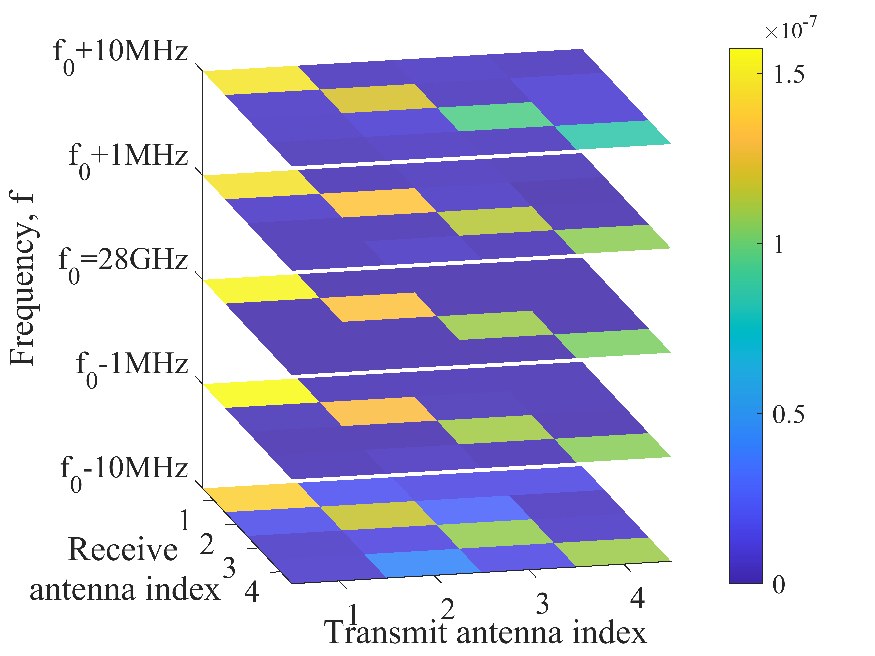}}
\caption{{The visualization of the end-to-end channel matrices $\mathbf{H}_i$ at different frequency points of SIM$_\text{SC}$ and SIM$_\text{MC}$ configurations.}}
    \label{visualization}
\end{figure}

Fig.~\ref{visualization} compares the end-to-end channel matrices $\mathbf{H}_i=\mathbf{Q}_i\mathbf{G}_i\mathbf{P}_i$ at different frequencies of SIM$_\text{SC}$ and SIM$_\text{MC}$ configurations, respectively. As shown in Fig.~\ref{fig:subfig1}, the SIM$_\text{SC}$ fails to form a channel diagonalization spanning from the transmit antennas to the receive antennas when the transmission frequency is not aligned with the center frequency $f_0$. This shows that SIM$_\text{SC}$ cannot maintain interference-free channel characteristics even within a narrow frequency range of 1 MHz, leading to significant IAI. In contrast, Fig.~\ref{fig:subfig2} demonstrates that the proposed SIM$_\text{MC}$ achieves more uniform and consistent diagonalization across a broader range within the effective bandwidth of 20 MHz. These results highlight the superiority of the proposed SIM$_\text{MC}$, where the SIM is configured accounting for multi-carriers within $B_e$. The improved adaptability of the proposed SIM$_\text{MC}$ in frequency-selective environments allows it to enhance system performance over a wider frequency spectrum, enabling more robust and interference-free multi-carrier transmission.

\subsection{Comparison with Existing Transmission Technologies}
~~To further demonstrate the performance superiority, the proposed wideband SIM$_\text{MC}$-enhanced MIMO OFDM system is benchmarked against a single-carrier optimized SIM$_\text{SC}$ in \cite{SIM-HMIMO} and a HMA/dynamic metasurface antennas (DMA)-aided MIMO system following the configuration in \cite{OFDM4}. To ensure a fair comparison, the DMA simulation follows the alternating optimization algorithm and hardware assumptions in \cite{OFDM4}, adjusted to match our SIM setup. The channel capacity performance of three representative architectures is compared under the same simulation parameters, including the number of antennas, subcarriers, system bandwidth, and total transmit power range.

\begin{figure}
	\centerline{\includegraphics[width=0.4\textwidth]{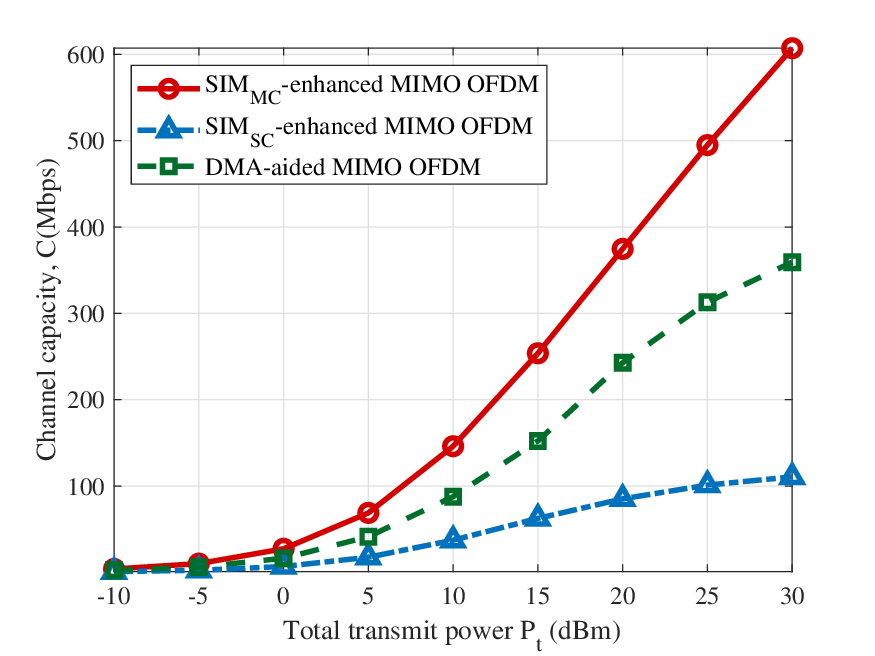}}
	\caption{{{The channel capacity $C$ of the proposed SIM$_\text{MC}$-enhanced, SIM$_\text{SC}$-enhanced, and DMA-aided MIMO OFDM systems under different total transmit powers $P_t$. }}}
	\label{DMACom} 
\end{figure}

Fig.~\ref{DMACom} demonstrates the superiority of our proposed SIM$_\text{MC}$ architecture. Across all transmit power levels, SIM$_\text{MC}$ achieves significantly higher capacity, especially in the high-power regime, where it outperforms DMA by over 100 Mbps and SIM$_\text{SC}$ by more than 200 Mbps. While the SIM$_\text{SC}$ approach suffers from performance saturation due to its inability to track frequency selectivity, the DMA-based system achieves moderate gains but remains constrained by its analog combining structure along microstrip waveguides and limited spatial diversity. Unlike the DMA, the proposed SIM$_\text{MC}$ allows subcarrier-dependent phase optimization across its multiple transmissive layers, with additional DoF, to compensate for frequency-selective fading and inter-carrier interference. Due to the fully-passive, stackable design of SIM$_\text{MC}$, it avoids the feedline loss and spatial constraints associated with DMA's microstrip-fed subwavelength resonators, making SIM more suitable for large-aperture, high-resolution beamforming. Furthermore, our SIM system achieves these gains under a fully-analog hardware structure, without requiring extra RF chains or ADCs at the metasurface. Compared to DMA, which requires one RF chain per microstrip and baseband digital processing, SIM$_\text{MC}$ is substantially more energy-efficient and hardware-friendly for mmWave and THz applications.

The proposed SIM$_\text{MC}$ is also compared against its single-layer metasurface counterpart, which can be seen as the fully-analog RIS configuration considering multiple subcarriers. To ensure fairness, the total number of meta-atoms in both the TX-SIM and the RX-SIM is reconfigured into a single-layer metasurface, maintaining the same overall transceiver metasurface area. To make the overall metasurface area consistent, the spacings between adjacent meta-atoms of the single-layer metasurface in the TX and RX are set to $(5c)/(f_0\left \lceil \sqrt{ML}  \right \rceil) $ and $(5c)/(f_0\left \lceil \sqrt{NK}  \right \rceil) $, respectively, and all other parameters are kept consistent. Note that the digital precoding and combining are removed in both schemes.

\begin{figure}
	\centerline{\includegraphics[width=0.4\textwidth]{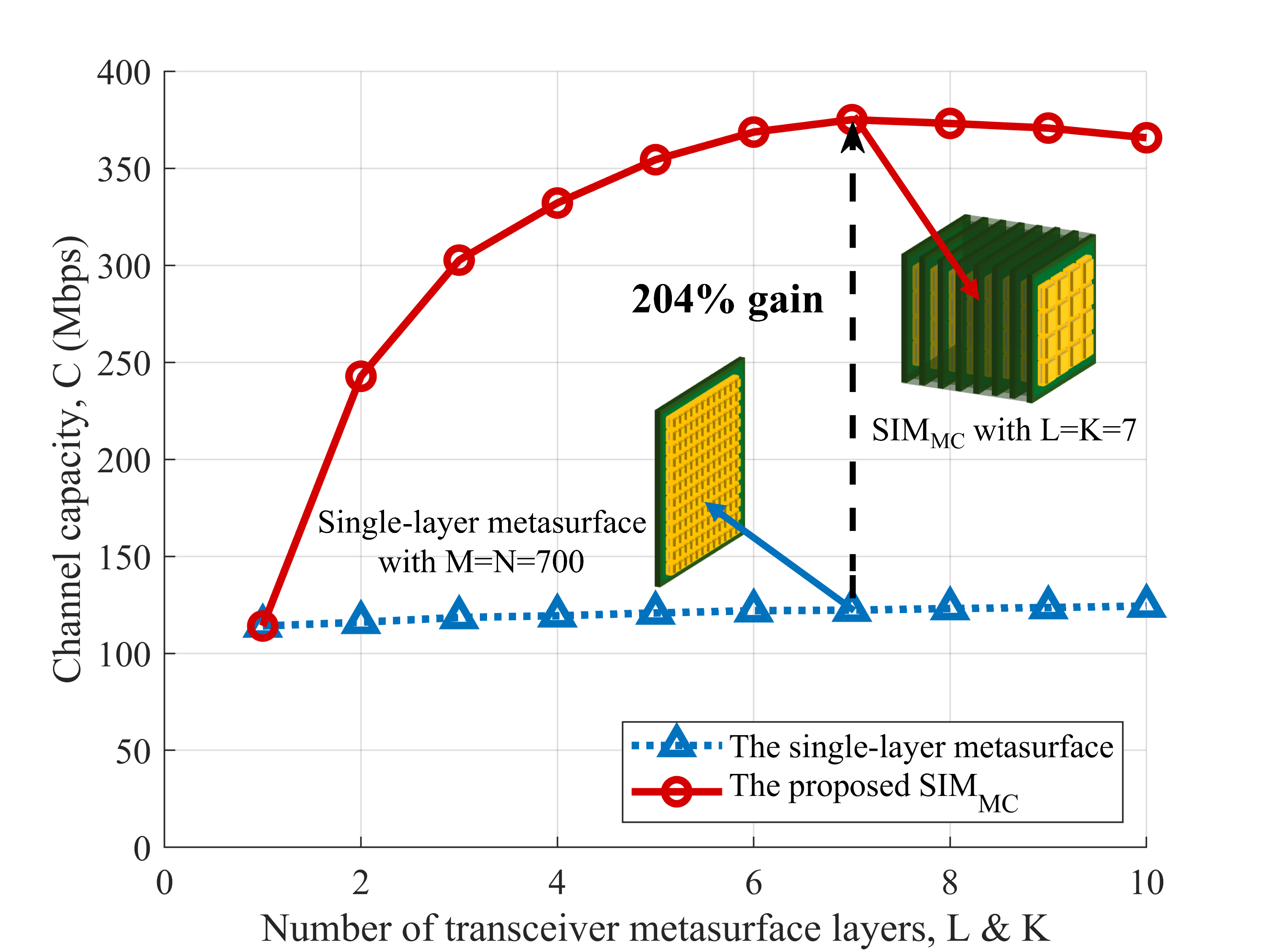}}
	\caption{{The channel capacity comparison versus the number of transceiver metasurface layers $L$ and $K$ of the proposed SIM$_\text{MC}$ and its single-layer metasurface counterpart that has the same total number of meta-atoms.}}
	\label{changeCom} 
\end{figure}

As shown in Fig.~\ref{changeCom}, the proposed SIM$_\text{MC}$ demonstrates a significant performance advantage over the single-layer metasurface in terms of channel capacity. When the number of the transceiver metasurface layers is $L=K=1$, the proposed SIM$_\text{MC}$ and its single-layer metasurface counterpart have the same value of channel capacity because they have the same structure, $M=N=100$ mate-atoms and 1 metasurface layer. However, the single-layer metasurface exhibits small capacity gains, even when the number of meta-atoms per layer is increased from 100 to 1000. The single-layer metasurface struggles to suppress IAI due to insufficient programmable DoF. Although the overall transceiver metasurface areas are the same, the channel capacity of the SIM$_\text{MC}$ becomes much higher than the single-layer metasurface as $L$ and $K$ increase, and the capacity of the SIM$_\text{MC}$ reaches a peak at $L=K=7$. Specifically, the SIM$_\text{MC}$ achieves a total channel capacity of 375 Mbps within the effective bandwidth, representing a 204\% improvement over the single-layer metasurface, which attains only 123 Mbps. These results highlight the advantages of the proposed SIM$_\text{MC}$ architecture in achieving robust, high-capacity MIMO OFDM communications.

\section{Conclusions}
\label{Section6}
~~This paper proposed a novel SIM-enhanced MIMO OFDM communication system, specifically designed to tackle the challenges of multi-carrier transmission in wideband channels. The proposed system introduces two key innovations: wave-domain fully-analog beamforming to effectively suppress IAI and programmable multi-path control to mitigate adverse deep fading caused by frequency-selective multi-path propagation, thereby enhancing OFDM signal transmission performance.

By employing the BCD-PCCP optimization algorithm, the SIM is configured to ensure interference-free transmission and achieve near-perfect diagonalization of the end-to-end channel within 20 MHz. The system's scalability is also validated by optimizing subcarrier spacing, the number of subcarriers, and structural parameters such as the number of metasurface layers and meta-atoms per layer. At a center frequency of 28 GHz with an effective bandwidth of 20~MHz over 16 subcarriers, \textbf{the capacity achieved by using the proposed SIM configuration is improved by more than 300\% compared to the conventional SIM configuration method that only accounts for the single frequency, over 60\% gain than the DMA-aided system, and over 200\% gain than the single-layer metasurface counterpart}. Furthermore, the results underscore the system's ability to adapt dynamically to frequency-selective multi-path fading while maintaining practical optimization complexity.

In summary, the fully-analog SIM-enhanced MIMO OFDM architecture removes partial digital baseband processing, significantly reducing energy consumption and hardware complexity. Further studies will extend SIM-enhanced MIMO OFDM architectures to multi-user scenarios, alternative wideband transmission schemes, and real-world implementation challenges, further advancing high-capacity and energy-efficient wireless communications.

\section*{Appendix A \\ Proof of Lemma 1}
~~For convenience, let $\mathrm{\Gamma}_{i}={||\alpha\mathbf{Q}_i\mathbf{G}_i\mathbf{P}_i-[\mathbf{\Lambda}_i]_{1:S,1:S}||}_F^2$. Then, expanding the Frobenius norm yields:
\begin{align}
\label{trace}
\mathrm{\Gamma}_{i} =&\alpha^2 {\rm Tr}(\mathbf{Q}_i\mathbf{G}_i\mathbf{P}_i\mathbf{P}_i^H\mathbf{G}_i^H\mathbf{Q}_i^H)-\alpha{\rm Tr}(\mathbf{Q}_i\mathbf{G}_i\mathbf{P}_i[\mathbf{\Lambda}_i]_{1:S,1:S}^H)-\nonumber\\
&\alpha^*{\rm Tr}([\mathbf{\Lambda}_i]_{1:S,1:S}\mathbf{P}_i^H\mathbf{G}_i^H\mathbf{Q}_i^H)-{\rm Tr}([\mathbf{\Lambda}_i]_{1:S,1:S}[\mathbf{\Lambda}_i]_{1:S,1:S}^H).
\end{align}

Decompose $\mathbf{P}_i = \mathbf{P}_i^\mathbb{L} \mathbf{\Phi}^l \mathbf{P}_i^\mathbb{R}$ and $\mathbf{Q}_i = \mathbf{Q}_i^\mathbb{L} \mathbf{\Psi}^k \mathbf{Q}_i^\mathbb{R}$. Using matrix vectorization~\cite{Vec}, the first term becomes:
\begin{align}
\label{firstP}
&{\rm Tr}(\mathbf{Q}_i\mathbf{G}_i\mathbf{P}_i\mathbf{P}_i^H\mathbf{G}_i^H\mathbf{Q}_i^H) \nonumber\\
&= {\rm Tr}(\mathbf{Q}_i\mathbf{G}_i\mathbf{P}^\mathbb{L}_i\mathbf{\Phi}^l\mathbf{P}^\mathbb{R}_i{\mathbf{P}^\mathbb{R}_i}^H{\mathbf{\Phi}^l}^H{\mathbf{P}^\mathbb{L}_i}^H{\mathbf{G}_i}^H{\mathbf{Q}_i}^H)  \nonumber\\
&={\rm Tr}(({\rm diag}({\bm{\phi}^l}))^H{\mathbf{P}^\mathbb{L}_i}^H{\mathbf{G}_i}^H{\mathbf{Q}_i}^H\mathbf{Q}_i\mathbf{G}_i\mathbf{P}^\mathbb{L}_i {\rm diag}({\bm{\phi}^l})\mathbf{P}^\mathbb{R}_i{\mathbf{P}^\mathbb{R}_i}^H) \nonumber\\
& = ({\bm{\phi}^l})^H({\mathbf{P}^\mathbb{L}_i}^H{\mathbf{G}_i}^H{\mathbf{Q}_i}^H\mathbf{Q}_i\mathbf{G}_i\mathbf{P}^\mathbb{L}_i)\odot(\mathbf{P}^\mathbb{R}_i{\mathbf{P}^\mathbb{R}_i}^H) {\bm{\phi}^l}.
\end{align}

Since ${\mathbf{P}^\mathbb{L}_i}^H{\mathbf{G}_i}^H{\mathbf{Q}_i}^H\mathbf{Q}_i\mathbf{G}_i\mathbf{P}^\mathbb{L}_i$ and $\mathbf{P}^\mathbb{R}_i{\mathbf{P}^\mathbb{R}_i}^H$ are Hermitian matrices, their Hadamard product is also a Hermitian matrix. 
\begin{align}
\label{firstQ}
&{\rm Tr}(\mathbf{Q}_i\mathbf{G}_i\mathbf{P}_i\mathbf{P}_i^H\mathbf{G}_i^H\mathbf{Q}_i^H)\nonumber\\
&= {\rm Tr}(\mathbf{Q}^\mathbb{L}_i\mathbf{\Psi}^k\mathbf{Q}^\mathbb{R}_i\mathbf{G}_i\mathbf{P}_i{\mathbf{P}_i}^H{\mathbf{G}_i}^H{\mathbf{Q}^\mathbb{R}_i}^H{\mathbf{\Psi}^k}^H{\mathbf{Q}^\mathbb{L}_i}^H)\nonumber\\
& = ({\bm{\psi}^k})^H({\mathbf{Q}^\mathbb{L}_i}^H\mathbf{Q}^\mathbb{L}_i) \odot (\mathbf{Q}^\mathbb{R}_i\mathbf{G}_i\mathbf{P}_i{\mathbf{P}_i}^H{\mathbf{G}_i}^H{\mathbf{Q}^\mathbb{R}_i}^H){\bm{\psi}^k}.
\end{align}

Similarly, the second and third terms are expressed as:
\begin{align}
&{\rm Tr}(\mathbf{Q}_i\mathbf{G}_i\mathbf{P}_i[\mathbf{\Lambda}_i]_{1:S,1:S}^H)={\rm Tr}(\mathbf{\Phi}^l\mathbf{P}^\mathbb{R}_i[\mathbf{\Lambda}_i]_{1:S,1:S}^H\mathbf{Q}_i\mathbf{G}_i\mathbf{P}^\mathbb{L}_i) \nonumber\\
&=({\rm vec}({\rm diag}({\bm{\phi}^l}))^H)^H {\rm vec}(\mathbf{P}^\mathbb{R}_i[\mathbf{\Lambda}_i]_{1:S,1:S}^H\mathbf{Q}_i\mathbf{G}_i\mathbf{P}^\mathbb{L}_i), \label{secondP} \\
&={\rm Tr}(\mathbf{Q}^\mathbb{L}_i\mathbf{\Psi}^k\mathbf{Q}^\mathbb{R}_i\mathbf{G}_i\mathbf{P}_i[\mathbf{\Lambda}_i]_{1:S,1:S}^H) \nonumber\\
& = ({\rm vec}({\rm diag}(\bm{\psi}^k))^H)^H {\rm vec}(\mathbf{Q}^\mathbb{R}_i\mathbf{G}_i\mathbf{P}_i[\mathbf{\Lambda}_i]_{1:S,1:S}^H\mathbf{Q}^\mathbb{L}_i), \label{secondQ}
\end{align}
\begin{align}
&{\rm Tr}([\mathbf{\Lambda}_i]_{1:S,1:S}\mathbf{P}_i^H\mathbf{G}_i^H\mathbf{Q}_i^H)= \nonumber\\
&= ({\rm vec}({\rm diag}({\bm{\phi}^l})))^H {\rm vec}({\mathbf{P}^\mathbb{L}_i}^H\mathbf{G}_i^H\mathbf{Q}_i^H[\mathbf{\Lambda}_i]_{1:S,1:S}{\mathbf{P}^\mathbb{R}_i}^H), \label{thirdP} \\
&= ({\rm vec}({\rm diag}({\bm{\psi}^k})))^H{\rm vec}({\mathbf{Q}^\mathbb{L}_i}^H[\mathbf{\Lambda}_i]_{1:S,1:S}\mathbf{P}_i^H\mathbf{G}_i^H{\mathbf{Q}^\mathbb{R}_i}^H). \label{thirdQ}
\end{align}

Finally, the fourth term ${\rm Tr}([\mathbf{\Lambda}_i]_{1:S,1:S}[\mathbf{\Lambda}_i]_{1:S,1:S}^H)$ is a constant. Thus, the term $\mathrm{\Gamma}_{i}$ forms a standard quadratic form w.r.t. $\bm{\phi}^l$ and $\bm{\psi}^k$, repsectively. Since the quadratic terms are convex and the linear terms do not affect the convexity, the objective function $\mathrm{\Gamma}=\sum_{i=1}^{N_c}\mathrm{\Gamma}_{i}$ is convex.

\end{document}